\documentclass[journal = mpohbp]{achemso}

\usepackage{graphicx}
\usepackage{xcolor}
\usepackage[draft]{fixme}

\usepackage[latin1]{inputenc}
\usepackage{amsmath}
\usepackage{latexsym}
\usepackage{float}
\usepackage{enumerate}
\usepackage{amsfonts}
\usepackage{amssymb}

\title{
Using cluster theory to calculate the experimental structure factors of antibody solutions
}

\author{Nicholas Skar-Gislinge}
\affiliation{Physical Chemistry, Department of Chemistry, Lund University, SE-221 00 Lund, Sweden}
\altaffiliation{present address: Copenhagen Business School, Porcelaenshaven 18B, 2000 Frederiksberg, Denmark}

\author{Fabrizio Camerin}
\affiliation{Soft Condensed Matter, Debye Institute for Nanomaterials Science, Utrecht University, Princetonplein 5, 3584 CC Utrecht, The Netherlands}

\author{Anna Stradner} 
\affiliation{Physical Chemistry, Department of Chemistry, Lund University, SE-221 00 Lund, Sweden}
\altaffiliation{LINXS - Lund Institute of advanced Neutron and X-ray Science, Scheelev{\"a}gen 19, SE-223 70 Lund, Sweden}

\author{Emanuela Zaccarelli} \email{emanuela.zaccarelli@cnr.it}
\affiliation{Institute for Complex Systems, National Research Council (ISC-CNR), Piazzale Aldo Moro 5, 00185 Rome, Italy}
\alsoaffiliation{Department of Physics, Sapienza University of Rome, Piazzale Aldo Moro 2, 00185 Rome, Italy}

\author{Peter Schurtenberger} \email{peter.schurtenberger@fkem1.lu.se}
\affiliation{Physical Chemistry, Department of Chemistry, Lund University, SE-221 00 Lund, Sweden}
\altaffiliation{LINXS - Lund Institute of advanced Neutron and X-ray Science, Scheelev{\"a}gen 19, SE-223 70 Lund, Sweden}


\begin{document}

\begin{abstract}

Monoclonal antibody solutions are set to become a major therapeutic tool in the years to come, capable of targeting various diseases by clever designing their antigen binding site. However, the formulation of stable solutions suitable for patient self-administration typically presents challenges, as a result of the increase in viscosity that often occurs at high concentrations. Here, we establish a link between the microscopic molecular details and the resulting properties of an antibody solution through the characterization of clusters, which arise in the presence of self-associating antibodies. In particular, we find that experimental small-angle X-ray scattering data can be interpreted by means of analytical models previously exploited for the study of polymeric and colloidal objects, based on the presence of such clusters. The latter are determined by theoretical calculations and supported by computer simulations of a coarse-grained minimal model, in which antibodies are treated as Y-shaped colloidal molecules and attractive domains are designed as patches. Using the theoretically-predicted cluster size distributions, we are able to describe the experimental structure factors over a wide range of concentration and salt conditions. We thus provide microscopic evidence for the well-established fact that the concentration-dependent increase in viscosity is originated by the presence of clusters. Our findings bring new insights on the self-assembly of monoclonal antibodies, which can be exploited for guiding the formulation of stable and effective antibody solutions.

\end{abstract}

\begin{tocentry}
\includegraphics[scale=0.25]{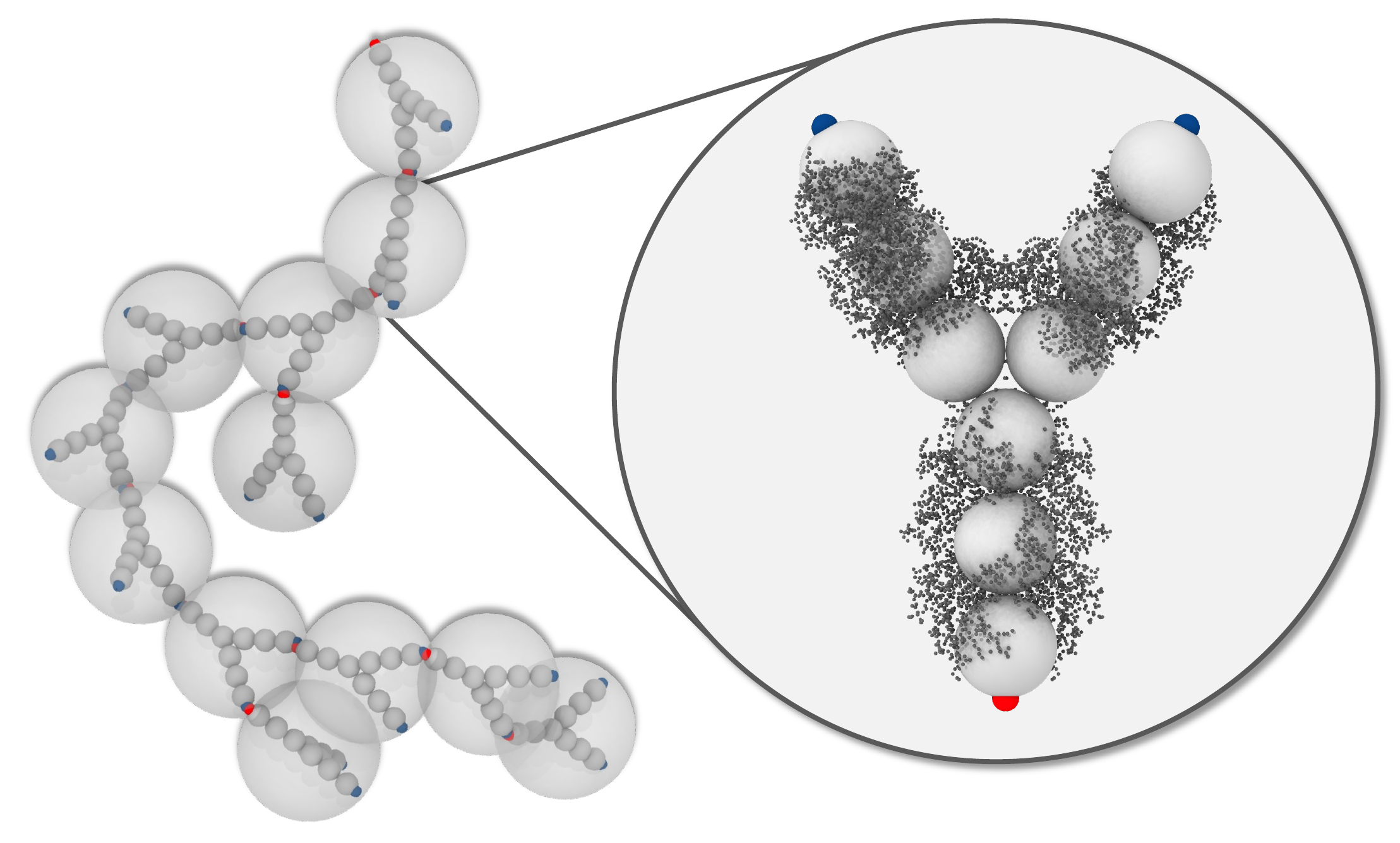}
\end{tocentry}

\maketitle

\section{Introduction}

Monoclonal antibodies (mAbs) have moved into the focus of pharmaceutical industry as a major platform for potential drug candidates \cite{Beck2010, Nelson2010, Reichert2012, Lu2020}. However, successful mAb applications that allow for facile home administration require stable and low viscosity high concentration formulations with concentrations of the order of $100$ g/l or more, which are often difficult to achieve \cite{Berteau2015, Bittner2018, Narasimhan2012,Shire2009}. In fact, mAbs are prone to exhibit reversible self-association at high concentrations that results in enhanced viscosity~\cite{liu2006reversible,Tomar2016,Wang2018}, which creates the need for an advanced predictive understanding of concentrated protein solutions. This is particularly important as a high solution viscosity is also a challenge for the production of high concentration stable biologics. 

In the protein literature, there is already significant evidence that the formation of (equilibrium) transient clusters strongly influences the relative viscosity~\cite{Cardinaux2011,Lilyestrom2013,wang2018cluster,Bergman2019}. This often results in the occurrence of  dynamic arrest through a so-called cluster glass transition, as long as the lifetime of the transient bonds between proteins or colloids is long enough~\cite{Cardinaux2011}. The presence of such clusters strongly influences the zero shear viscosity $\eta_0$ of concentrated solutions, resulting in an arrest transition at lower concentrations when compared to a purely monomeric solution~\cite{Bergman2019}. Specifically to mAbs, there are several studies showing that the increased viscosity in concentrated solutions of mAbs is linked to cluster formation \cite{VonBulow2019, Yearley2014, Lilyestrom2013, Chowdhury2020, Buck2015, Godfrin2016, Skar-Gislinge2019,dear2019x,wang2018cluster,ausserwoger2022non,lai2021calculation,chowdhury2023characterizing}. Many of these works have made attempts to characterize cluster formation in mAb solutions, and to interpret antibody solution properties through analogies with colloids or polymers. In particular, experimental scattering techniques were used to investigate protein interactions and self-association~\cite{Yearley2013, Yearley2014,Scherer2013,Castellanos2014,Godfrin2016,Corbett2017,calero2016coarse}. Numerically, the first coarse-grained model for the study of antibody self-association dates back to 2012, where Chaudhri and coworkers proposed a 12 and 26 bead-based models arranged in a Y-shape and demonstrated the formation of clusters for two model antibodies~\cite{chaudhri2012coarse}. Later, several other works have laid the foundations for antibody models that explicitly include charged domains~\cite{chaudhri2013role,tomar2016molecular,izadi2020multiscale,wang2018structure}. However, we are still far from having any predictive understanding and a generally accepted methodology and/or theoretical framework to detect antibody cluster formation.
The transient antibody clusters are formed by large and flexible molecules interacting through a number of different intermolecular forces. This makes a theoretical treatment providing a quantitative link between the molecular structure, intermolecular interactions and experimentally obtained dynamic quantities very challenging.
In particular, at high concentrations, it requires different coarse-graining strategies, able to incorporate crucial molecular information into colloid-like models that are then amenable to computer simulations as well as to analytical calculations. 

The aim of the present work is to explore the application of small-angle X-ray scattering (SAXS) experiments and investigate whether these are able to provide us with a typical "fingerprint" for the presence of small equilibrium clusters. We focus on a well-characterized model system of a humanized IgG4 against trinitrophenyl, which was found to exhibit an increased viscosity at high concentrations \cite{Neergaard2013}. We have previously reported a detailed experimental investigation of key structural and dynamic properties~\cite{Skar-Gislinge2019}, that were rationalized through a relatively simple theoretical framework. In particular,  
we proposed a colloid-inspired coarse-grained approach where we explicitly considered the anisotropy of both shape and interactions of the antibody molecules. 
We focused on a simple patchy model, that is built from calculations of the electrostatic properties of the considered mAbs~\cite{Skar-Gislinge2019}, condensing the long-range interactions into specific attractive sites.
Such model retains the minimal ingredients to describe correctly the antibody self-association and has the advantage to be analytically treatable with Wertheim~\cite{Wertheim1984} and hyperbranched polymer theories~\cite{Rubinstein2003}.
This allowed us to predict the cluster size distribution as a function of antibody concentration, thus being able to successfully reproduce the experimental data. However, in our previous work, the presence of self-assembled clusters was derived only indirectly from \textit{macroscopic} experimental quantities, namely the apparent weight-average molecular weight $\big\langle M_{w} \big\rangle_{\mathrm{app}}$ obtained by static light scattering (SLS), the apparent $z$-average hydrodynamic radius $\big\langle R_{h} \big\rangle_{\mathrm{app}}$ measured by dynamic light scattering (DLS) and the relative zero shear viscosity $\eta_r = \eta_0 / \eta_s$ where $\eta_0$ is the zero shear viscosity of the antibody solution and $\eta_s$ the solvent viscosity, obtained by microrheology~\cite{Skar-Gislinge2019}.

Here, we aim to provide a \textit{microscopic} evidence of cluster formation by including in our analysis new SAXS measurements for the same antibodies at different salt concentrations. The use of this technique provides high resolution structural data down to the molecular scale. In turn, such data are used to derive an analytical model for the scattering signal of antibody clusters as a function of the cluster size or aggregation number $s$, based on cluster theory for polymeric and colloidal objects. 
We further test the theoretical model against the results of computer simulations, where we improve our minimal model for Y-shaped antibodies put forward in Ref.~\cite{Skar-Gislinge2019}. 
The additional step in the theoretical treatment performed in this work finally
leads to the analytical calculation of the experimentally measured structure factors at all wavevectors for different mAb concentrations.
This novel finding
thus constitutes the main result of the present work.
Our calculations are found to well reproduce the experimental data from SAXS, providing a clear microscopic signature of the presence of small clusters in antibody solutions.

\section{Methods}
\small
\subsection{Experimental sample preparation}

In this study we have used a humanized IgG4 antibody against trinitrophenyl (TNP). The antibody was manufactured by Novo Nordisk A/S and purified using Protein A chromatography, and subsequently concentrated and buffer exchanged into a 10mM Histidine, 10mM NaCl, pH 6.5 buffer at a concentration of 100mg/mL. From this stock solution, samples were prepared by concentrating and buffer exchanging into either 20mM Histidine, 10mM NaCl, pH 6.5 or 20mM Histidine, 50mM NaCl, pH 6.5 buffers using Amicon spin filters with a molecular weight cutoff at 100kD (Merck, Germany). The two different solvents thus have a total ionic strength of either 17 mM or 57 mM, respectively.
Samples with decreasing concentration were then prepared from the concentrated sample by dilution, determining the antibody concentration using UV/Vis absorption at 280nm and a molecular extinction coefficient derived from the amino acid composition of 223400 M$^{-1}$ cm$^{-1}$ ($E^{280nm}_{0.1\%, 1cm}$ = 1.489 g$^{-1}$ L cm$^{-1}$).

\subsection{Light scattering measurements}

SLS experiments were performed using a 3D-LS Spectrometer (LS Instruments AG, Switzerland) with a 632nm laser, recording DLS and SLS data simultaneously. The measurements were conducted at $90^{\circ}$ scattering angle. Before measurement, the samples were transferred to pre-cleaned 5mm NMR tubes  and centrifuged at 3000 g and 25 $^{\circ}$C for 15 min, to remove any large particles and to equilibrate temperature. Directly after centrifugation, the samples were placed in the temperature equilibrated sample vat and the measurement was started after 5 minutes to allow for thermal equilibration. Additional low concentration SLS measurements were done using a HELIOS DAWN multi-angle light scattering instrument (Wyatt Technology Corporation, CA, USA), connected to a concentration gradient pump. The instruments were calibrated to absolute scale using toluene (with a Rayleigh ratio of  $1.37\cdot 10^{-5}$ cm$^{-1}$ at $25^{\circ}$ C and $\lambda = 632.8$ nm) in the case of the 3D-LS Spectrometer, and toluene and a secondary protein standard with a known molecular mass for the HELIOS DAWN, allowing for direct comparison of the two data sets.
	
From the SLS experiments, the apparent weight average molecular weight $\langle M \rangle_{w,app}$ of the antibodies in solution was calculated using
	
\begin{equation}\label{SLS1}
\langle M \rangle_{w,app} = \frac{R(90)}{K C}
\end{equation}

\noindent where $R(90)$ is the absolute excess scattering intensity or excess Rayleigh ratio measured at a scattering angle of $90^{\circ}$, $K = 4 \pi^2 n^2 ( d n / d C )^2 /N_A \lambda^4_0$, $n$ is the refractive index of the solution, $dn/dC = 0.192$ L/g  is the refractive index increment of the antibodies, $\lambda_0$ is the vacuum wavelength of the laser, and $C$ is the antibody concentration in milligrams per milliliter. The excess Rayleigh ratio $R(90)$ is obtained from the measured scattering of the protein solution, $I(90)$, of the solvent $I_s(90)$ and of the reference standard $I_{ref}(90)$ using $R(90) = [(I(90) - I_s(90))/I_{ref}(90)] R_{ref} (n/n_{ref})^2$, where $R_{ref}$ is the Rayleigh ratio of the reference solvent, and $n$ and $n_{ref}$ are the index of refraction of the solution and the reference solvent, respectively. Note that, due to the small size of the antibody molecules and of the antibody clusters, there is no measurable angular dependence in the scattering intensity, and we can directly use the intensity values measured at a scattering angle of $90^{\circ}$ instead of the corresponding values extrapolated to $\theta = 0$. 

\subsection{Microrheology}
The zero shear viscosity $\eta_0$ of the antibody solutions relative to that of the pure buffer, denoted as the relative viscosity $\eta_r = \eta_0/\eta_s$, was obtained using DLS-based tracer microrheology. Sterically stabilized (pegylated) latex particles were mixed with protein samples to a concentration of 0.01 $\% $v/v using vortexing and transferred to 5 mm NMR tubes.
	
The sterically stabilized particles were prepared by cross-linking 0.75 kDa amine-PEG (poly-ethylene glycol) (Rapp Polymere, 12750-2) to carboxylate stabilized polystyrene (PS) particles (ThermoFischer Scientific, C37483) with a diameter of 1.0 $\mu$m using EDC (N-(3-Dimethylaminopropyl)-N'-ethylcarbodiimide) (Sigma Aldrich, 39391) as described in detail in Ref.\cite{Garting2018}.
	
DLS measurements were performed on a 3D-LS Spectrometer (LS Instruments AG, Switzerland) at a scattering angle of 46-50$^\circ$ to avoid the particle form factor minima and thus maximise the scattering contribution from the tracer particles with respect to the protein scattering. Measurements were made using modulated 3D cross correlation DLS \cite{Block2010} to suppress all contributions from multiple scattering that occur, in the attempt to achieve conditions where the total scattering intensity is dominated by the contribution from the tracer particles. Samples were either prepared individually or diluted from more concentrated samples using a particle dispersion with the same particle concentration as in the sample as the diluent. The diffusion coefficient $D$ of the particles was then extracted from the intensity autocorrelation function using a first order cumulant analysis of the relevant decay. This diffusion coefficient is compared to that of the tracer particles in a protein-free solvent (buffer) resulting in a relative diffusion coefficient, $D_r = D_{Sample}/D_{Solvent}$, where $D_{Sample}$ is the measured diffusion coefficient of the tracer particles in the sample and $D_{Solvent}$ is the measured diffusion coefficient of the tracer particles in the pure solvent. For spherical particles with known hydrodynamic radius $R_H$ in the absence of measurable interparticle interaction effects the zero shear viscosity $\eta_i$ is related to the measured diffusion coefficient $D_i$ according to the Stokes-Einstein equation $D_{i}=\frac{k_BT}{6 \pi \eta_i R_H}$, where $i$ stands either for sample or solvent. 
Therefore the relative viscosity $\eta_r = \eta_{sample} / \eta_{solvent}$ is related to $D_r$ through $\eta_r = 1/D_r$ \cite{Garting2018,Furst2017}.

\subsection{SAXS measurements}

\subsubsection{Form factor}
In order to avoid any concentration-induced antibody clusters and other aggregates, the SAXS form factor of the  mAb was measured at the SWING beamline at synchrotron SOLEIL (Gif-sur-Yvette, France) using a combined size exclusion chromatography and SAXS setup. The setup consisted of an Agilant HPLC system composed of a BioSEC-3 300 column, an automatic sample loader and a UV/VIS detector, connected to a flow through cell located at the sample position in the SAXS instrument \cite{david_2009}. The sample loading and flow was controlled by the HPLC software, whereas the SAXS measurements were initiated manually. The SAXS measurements consisted of a background measurement $100 \times 1500$ $\mu s$ exposures once a stable UV/VIS baseline signal was acquired, and a sample measurement of $150 \times 1500$ $\mu s$ exposures initiated in order to cover the chromatogram. Between each exposure, a pause of $500$ $\mu s$ was automatically inserted in order to let the exposed material flow out of the exposed volume to minimize radiation damage. The azimuthal averaging of the detector image and absolute calibration of each frame was performed by the FOXTROT software available at the beamline, which also allowed for background subtraction, calculation of $R_g$, and forward scattering. After the data treatment, the scattered intensity was given in absolute units as a function of the scattering vector from $q = 0.00626$ \mbox{\normalfont\AA} $^{-1}$ to $q = 0.591$ \mbox{\normalfont\AA}$^{-1}$. The final scattering curve was composed by averaging the measurements around the central peak of the chromatogram. The concentration was determined using the UV/VIS absorption in the same area of the chromatogram measured by the HPLC UV/VIS detector. The time delay and peak broadening between the UV detector was determined using a protein standard.   

\subsubsection{Structure factors}
The higher concentration samples used to obtain the SAXS structure factors were measured on a pinhole camera (Ganesha 300 XL, SAXSLAB) covering a $q$-range from $0.003$ to $2.5$ \mbox{\normalfont\AA}$^{-1}$. In order to calculate the structure factors from data measured on two different SAXS instruments, the measured intensity data needed to be converted to the same scale and $q$ values. Common $q$ values were obtained by interpolating the measured intensities of the pinhole camera at the $q$ values of the SOLEIL data. In order to bring the samples measured on the pinhole camera to absolute scale, low concentration samples ($c \approx 3.8$ mg/ml) were measured for both ionic strengths on the Ganesha instrument, and a scaling factor maximizing the overlap between the measurements from the pinhole camera and the SOLEIL data at $q > 0.04$ \mbox{\normalfont\AA}$^{-1}$, where the contributions from the structure factor are negligible for these low concentrations, was determined. Using the interpolated intensities and this scaling factor, the structure factors were calculated by dividing the concentration normalized and re-scaled intensities $I(q)/c$ with those of the dilute sample measured at Soleil.

\subsection{Antibody model} 
In order to study the antibody collective behavior, each antibody is modeled in a coarse-grained fashion using a colloid-inspired approach. In particular, it consists of 9 beads arranged in a Y-shaped symmetric colloidal molecule, where each sphere has a unit-length diameter $\sigma$. The three central beads are arranged in an equilateral triangle, and the three arms of the Y, each made of three spheres, form angles of $150^{\circ}$ and $60^{\circ}$ with each other, see Figure~\ref{fig:model}(b). The geometric construction of the antibody implies that the circle tangent to the external sphere has a diameter $d_Y \approx 6.16\sigma$. 
Each bead in the coarse-grained Y model is a hard sphere with infinite repulsive potential at contact and each antibody is treated as a rigid body. 
The specific choice of a 9-bead model is justified by matching its excluded volume interactions to that of the hard sphere model system on which the patchy hard sphere model introduced below is based on, i.e. by calculating its excluded volume for different densities and by comparing it to the theoretical Carnahan-Starling prediction for hard spheres.
This aspect is discussed in-depth in the Supplementary Information file.
\\To account in a primitive fashion for the electrostatic-driven aggregation of the antibodies, the extremities of the three arms are decorated with patches of size $0.2633 \sigma$, one of type A on the tail and two of type B in the upper arms of the Y. This patch width allows to match in the simulations the bond probability $p$ determined from Wertheim theory, as described in more detail in a later section, although slightly exceeding the one-bond-per-patch condition that is assumed within the theory. However, we verified that the overall number of double bonds in simulations never exceeds a small percentage of the total for all considered mAb concentrations, thus allowing it to be safely ignored.
Bonds are allowed to occur only between A and B type patches and are modeled with an attractive square well potential of depth $\epsilon_0$, which sets the energy scale. AA and BB interactions are not considered.
The comparison between the radius of gyration of the experimental antibody and the Y-shaped model allows us to convert simulation units into real ones: being the former $R_g^{mAb} = 4.7$ nm and the latter $R_g^{hY} = 1.7297 \sigma$, we obtain the size of each bead in the model as $\sigma = 2.72$ nm.

\subsubsection{Monte Carlo simulations} 
We run Monte Carlo (MC) simulations with $N=1000$ Y-shaped antibodies. 
We start by preparing ten independent random configurations at each number density $\rho=N/V$, with $V$ the volume of the cubic simulation box. Then, we perform simulations at the desired temperature $T$ and average the results over the different configurations in order to improve statistics particularly at small scattering vectors. 
To perform simulations and analytical calculations at the same concentration as in experiments, we consider that the mass of a mAb molecule is 150 kDa. Therefore, at a weight concentration of 1 mg/ml, we have $4.098 \times 10^{15}$ particles/ml. With $\sigma^3 = 20.06$ nm$^3$, we obtain 1 ml $= 10^{21}$ nm$^3$ = $4.098 \times 10^{19} \sigma^3$. In this way, a weight concentration of 1 mg/ml or a particle number density of $4.098 \times 10^{15}$ particles/ml can be rewritten as $8.229 \times 10^{-5}$ particles/$\sigma^3$. 

\normalsize
\section{Results and discussion}

\subsection{Apparent aggregation number and relative viscosity}

\begin{figure}[t!]
\centering
\includegraphics[width=1\linewidth]{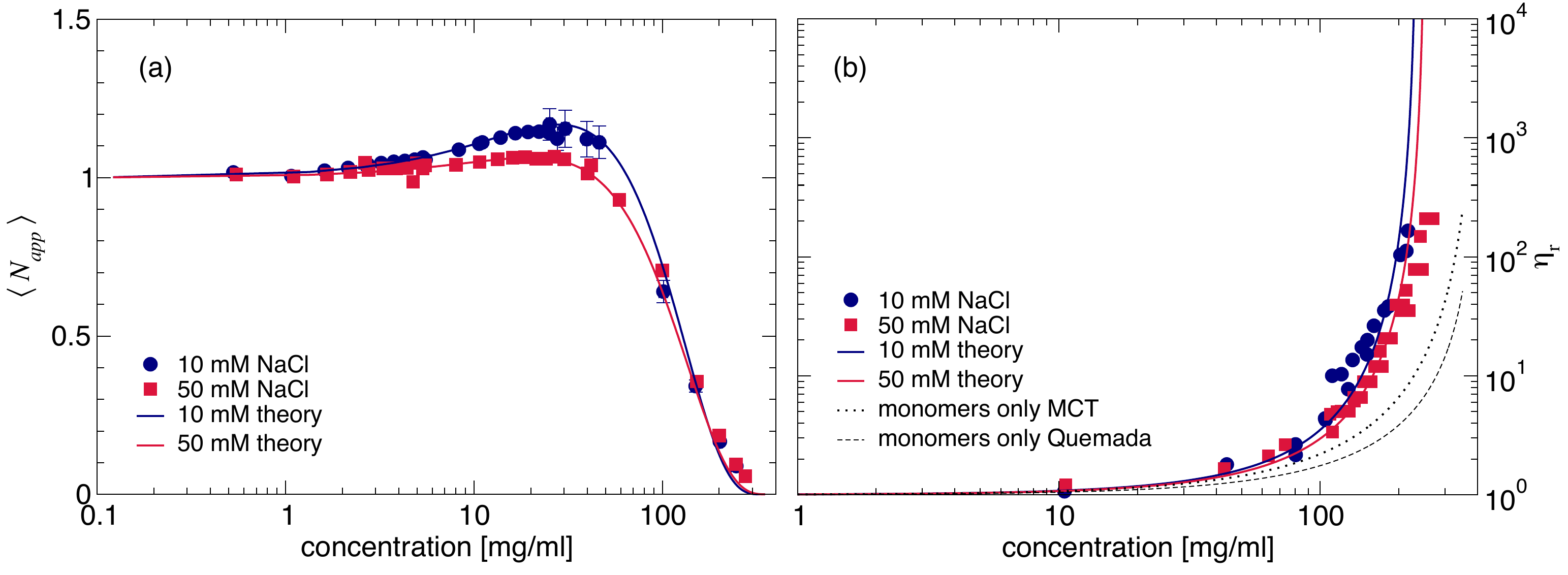}
 \caption{\footnotesize SLS  and microrheology data. (a) Experimental $\left\langle N_{app} \right\rangle$ as a function of $c$ for 10 mM NaCl (blue, 17 mM ionic strength) and 50 mM (red, 57 mM ionic strength) NaCl added, respectively. Also shown is a comparison with theoretical calculations (solid lines) based on a sticky hard sphere cluster model, see Eqns.~(\ref{naggapp_col}-\ref{lambda}). (b) Experimental $\eta_{r}$ as a function of $c$ for 10 mM NaCl (blue) and 50 mM (red) NaCl, respectively, together with the corresponding theoretical calculations (solid lines) from Eq.~\ref{powerlaw}) where $\phi_{HS}$ is calculated from eq.~\ref{phiHS}, $\gamma =3.0$ and $\phi_g=0.63$. A comparison with predictions for monomers only where the relative viscosity is also given either by using MCT (power law with exponent $\gamma=2.8$, dotted black line) or by the Quemada relationship for hard spheres (dashed black line)}.
	\label{Fig1-Napp-etar}
\end{figure}

Figure \ref{Fig1-Napp-etar} summarizes the concentration dependence of the key structural and dynamic quantities, namely the apparent aggregation number $\big\langle N_{app} \big\rangle$ and the relative viscosity $\eta_r$. Here $\big\langle N_{app} \big\rangle = \big\langle M_{w} \big\rangle_{\mathrm{app}}/M_1$, where $\big\langle M_{w} \big\rangle_{\mathrm{app}}$ is the apparent weight-average molecular weight measured by SLS and $M_1$ the molecular weight of the individual mAb, and $\eta_r = \eta_0 / \eta_s$, where $\eta_0$ is the zero shear viscosity of the mAb solution and $\eta_s$ is the solvent viscosity. The data shown in Fig.~\ref{Fig1-Napp-etar} is for two different total ionic strengths of the solvent, 17 mM (10 mM NaCl added to 20 mM Histidine buffer) and 57 mM (50 mM NaCl added to 20 mM Histidine buffer).
Both $\big\langle N_{app} \big\rangle$ and $\eta_r$ exhibit a behavior frequently found for proteins undergoing the formation of equilibrium clusters with a cluster size that increases with increasing concentration $c$~\cite{Lilyestrom2013, wang2018cluster}. For $\big\langle N_{app} \big\rangle$ this results in a non-monotonic concentration dependence with an initial weak increase due to the concentration-dependent cluster growth, followed by a strong decrease at higher values of $c$ due to the contributions from excluded volume interactions between clusters that become dominant at high concentrations~\cite{Skar-Gislinge2019}. In contrast, $\eta_r$ increases with increasing concentration and appears to diverge at a concentration of around 200-300 mg/ml, where the solution undergoes an arrest transition~\cite{Skar-Gislinge2019}.

In general, increasing the ionic strength in mAb solutions results in an enhanced propensity for self-assembly and cluster formation since stabilising charges on mAbs are screened, leading to a reduced electrostatic repulsion and thus colloidal stability \cite{Lilyestrom2013, Roberts2014a}. However, here we observe an opposite behavior, where the addition of salt actually reduces self-assembly, as evident from both experimental quantities. Such behavior is well known for proteins with oppositely charged patches \cite{Li2015} as also found for mAbs \cite{Liu2005}. As discussed below in more detail, under these conditions self-assembly is strongly influenced by the electrostatic attraction between oppositely charged patches despite an overall positive or negative charge, which is effectively screened by the addition of a large amount of salt.

\begin{figure}[t!]
\centering
\includegraphics[width=1\linewidth]{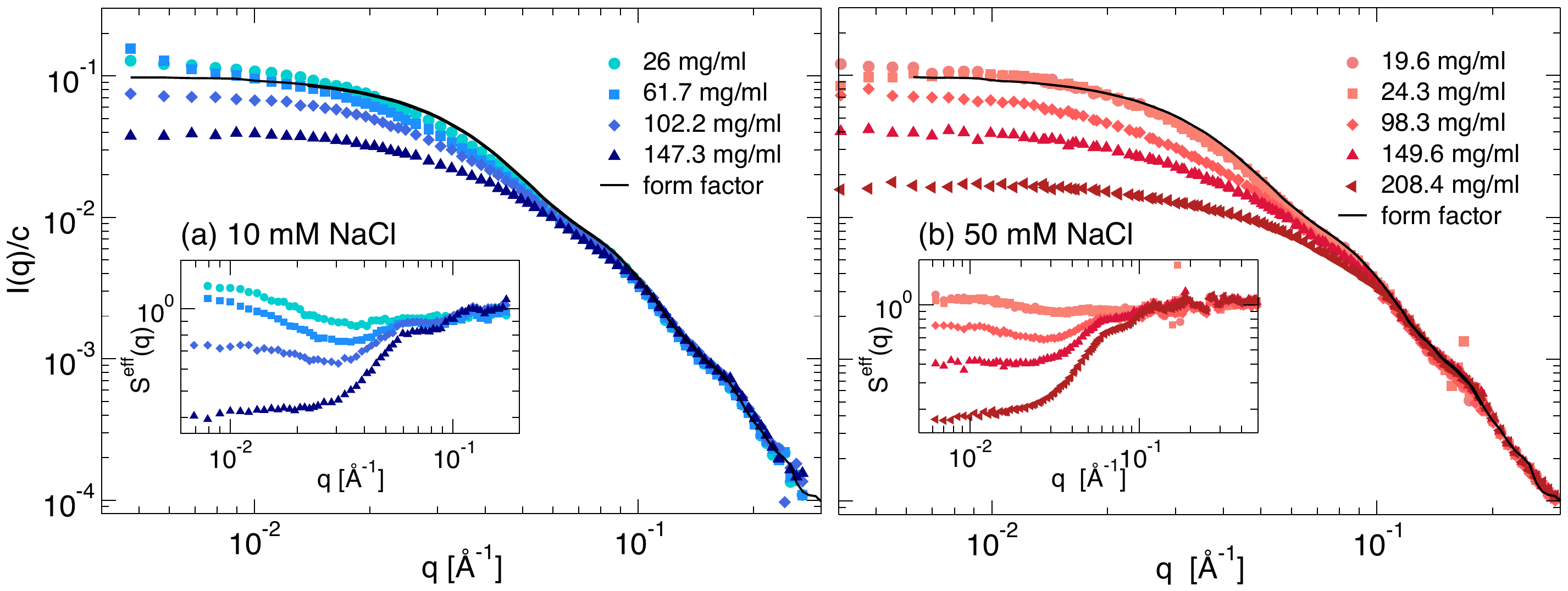}
  \caption{\footnotesize SAXS data for different mAb concentrations and ionic strengths. Experimental $I(q)$ as a function of $q$ for (a) 10 mM and (b) 50 mM NaCl added.
  Insets show the corresponding measured structure factors $S^{eff}(q)$ for both solvents.
  }
	\label{Fig-Iq}
\end{figure}

A similar pattern can also be observed from the results of the small-angle X-ray scattering (SAXS) experiments summarised in Fig. \ref{Fig-Iq}, which shows the concentration-normalized scattering data $I(q)/c$ as a function of the magnitude of the scattering vector $q$ for different mAb concentrations $c$ at both ionic strengths. We see the same non-monotonic $c$-dependence of the forward scattering as already observed for the SLS data shown in Fig. \ref{Fig1-Napp-etar}(a), while the high-$q$ data appears completely independent of concentration, indicating that the solution structure of the individual mAbs does not change with increasing concentration. Moreover, the initial increase of the forward intensity appears more pronounced at low ionic strength, while the data at higher concentrations and high ionic strength indicate a much more repulsive behavior characterised by a significant decrease of the data at low $q$-values. This is further illustrated in the insets of Fig. \ref{Fig-Iq} where we report plots of the effective or measured static structure factors $S^{eff}(q) = [I(q)/c]/[I_{ff}(q)/c_{ff}]$ for all data sets, where $I(q)$ and $I_{ff}(q)$ are the scattered intensities measured for the concentrated and the dilute (form factor) samples, and $c$ and $c_{ff}$ their concentrations, respectively. $S^{eff}(q)$ describes the influence of structural correlations only without the additional contributions from the monomer solution structure that is identical for all concentrations. Here, we use the notation $S^{eff}(q)$ to distinguish the effective structure factor measured in a static scattering experiment from the traditional structure factor $S(q) = \frac{1}{N} \sum_j \sum_k \left\langle  e^{-i {\bf q} \cdot ({\bf r_{j}-r_{k}})}\right\rangle $ defined in statistical physics, where $N$ is the number of particles and $\bf{r_{j,k}}$ refers to the position of the center of mass of particle j,k. While for monodisperse spherical particles $S^{eff}(q)$ and $S(q)$ are identical, this is not the case for anisotropic and/or polydisperse particles.\cite{Greene2016}

\subsection{Coarse-grained colloid models}

\begin{figure}[b!]
	\centering
        \includegraphics[width=1\linewidth]{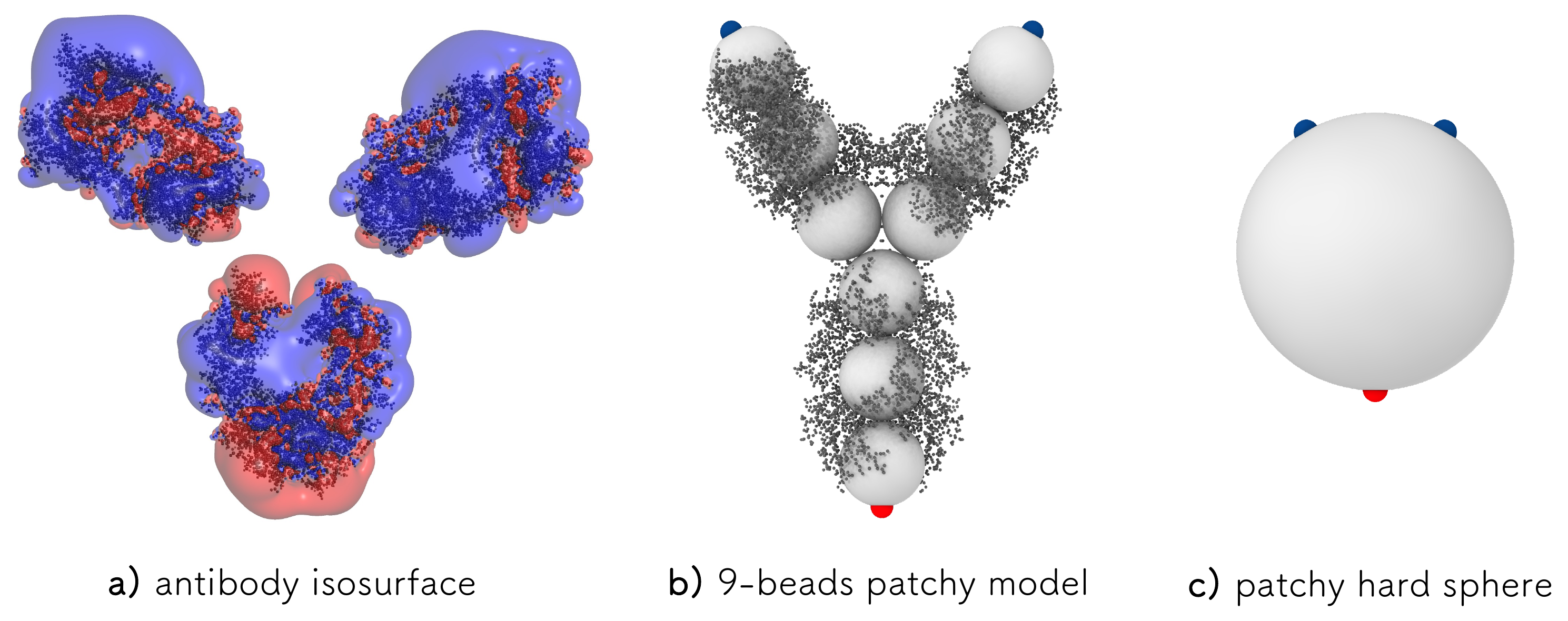}
	\caption{\footnotesize Design of the patchy model of mAbs. (a) Isosurfaces of the -1 (red) and +1 $k_BT$ (blue) electrostatic potential at pH $6.5$ with $10$ $mM$ NaCl, indicating an overall positive charge for the arms (FAB domains) and a largely negative charge for the tail (FC domain). Also shown is the atomistic representation of the antibody superimposed onto the isosurface potential. (b) Simulation snapshot of the YAB model: 9 hard spheres each of diameter $\sigma$ are constrained to a rigid Y shape, constituting a single mAb molecule. Each molecule is decorated with one $A$ patch on the tail (red) and two $B$ (blue) patches, one on each arm, mimicking the negative and positive charges respectively. Only $AB$ attractive interactions are considered mimicking the arm-to-tail electrostatic interactions. Furthermore, we also show the atomistic representation of the antibody. (c) Representation of the YAB model as an effective patchy hard sphere of diameter $\sigma_{HS}$ as in the Wertheim theory.
	}
	\label{fig:model}
\end{figure}

We have previously pointed out the importance of electrostatic interactions for the solution behavior of this mAb, and subsequently conducted a detailed study of the electrostatic isosurface of a single antibody molecule in the considered buffer solution~\cite{Skar-Gislinge2019}. The resulting charge distribution is illustrated in Fig.~\ref{fig:model}(a), which clearly shows that the considered mAbs have an overall positively charged surface on the two arms (FAB domains) and a largely negative charge on the tail (FC domain). This suggests that the main mechanism for aggregation of this particular mAb is an electrostatically driven attractive head-to-tail interaction, similarly to previous studies~\cite{chaudhri2013role}. Building on this hypothesis, we have thus operated a coarse-graining strategy based on a patchy colloid model that was capable to quantitatively reproduce the experimental findings for the lower ionic strength data set described in our earlier work~\cite{Skar-Gislinge2019}. The approach is illustrated in Fig.~\ref{fig:model}(b-c), where we include the 9-bead patchy model (YAB) used for computer simulations (Fig.~\ref{fig:model}(b)) and the patchy hard sphere model required for the analytical/numerical analysis (Fig.~\ref{fig:model}(c)).
Further modeling details are provided in Methods. 

Here, we first focus on the analysis of the experimental SLS data using a combination of Wertheim theory (WT) for patchy particles and hyperbranched polymer theory (hpt) that allows us to calculate the concentration dependence of the cluster size distribution compatible with the SLS data, and investigate whether the ionic strength dependence observed is also compatible with this approach. In Wertheim theory\cite{Wertheim1984,Tavares2010}, which is a thermodynamic perturbation theory, the YAB molecule is represented as an effective patchy sphere, illustrated in Fig. \ref{fig:model}(c), with a hard sphere diameter $\sigma_{HS}$. The free energy $F$ of a system of $N$ patchy spheres in a volume $V$, with number density $\rho=N/V$, is calculated as  the sum of the free energies of a hard sphere (HS) reference term $F_{HS}$ plus a bonding term $F_b$. For the reference term $F_{HS}$ we use the Carnahan-Starling (CS) free energy\cite{Hansennew} of an equivalent HS system, that has to be determined according to the nature of the molecule.  For non-spherical molecules, the HS reference system effective diameter $\sigma_{HS}$ is not known and needs to correctly take into account the excluded volume of the particles, which is established from the direct comparison to the experimental SLS data. The bonding free energy $F_b$ per particle of our 3-patch YAB model is then calculated as a function of the strength of the attraction $\epsilon_0$ and the bond probability $p_B$ as described in detail in Refs.~\cite{Bianchi2006, Skar-Gislinge2019}.

We can now perform a direct comparison between the analytical results for the YAB patchy model from WT and the experimental results for the apparent aggregation number $\big\langle N_{app} \big\rangle$  shown in Fig.~\ref{Fig1-Napp-etar}(a). To this aim, we calculate the isothermal compressibility $\kappa_T$ for our YAB model as a function of concentration by simply differentiating twice the analytic free energy $F$ with respect to volume~\cite{Tavares2009}, that is $\kappa_T=-\frac{1}{V} \left(\frac{\partial V}{\partial P}\right)_T= \frac{1}{V} \left(\frac{\partial^2 V}{\partial^2 F}\right)_T$ where $P$ is the pressure and $V$ is
the volume. Since $\kappa_T$ is related to $\big\langle N_{app} \big\rangle$ through

\begin{equation}\label{S0}
	\left\langle N_{app} \right\rangle = S^{eff}(0) \simeq \rho k_B T\kappa_T, 
\end{equation}

\noindent where $\rho$ is the number density of particles, Eq. \ref{S0} provides the link for the comparison between WT and the experimental results from SLS. 
By appropriately converting analytical and experimental results as described in Methods and in the SI, for the samples with 10 mM NaCl added we obtain good agreement between the experimentally measured $S^{eff}(0)$ and the Wertheim calculation  for $\sigma_{HS}=2.95\sigma$ and a temperature $T = 0.11$, which corresponds to a strength of the attraction between AB patches of $\epsilon_0 = 9.09$ $k_BT$. For the data at higher ionic strength, the attraction between the oppositely charged ends is slightly reduced due to the stronger screening, and we obtain best agreement for a temperature $T = 0.114$, which corresponds to a strength of the attraction between AB patches of $\epsilon_0 = 8.77$ $k_BT$.

\begin{figure}[t!]
	\centering
	\includegraphics[width=1\linewidth]{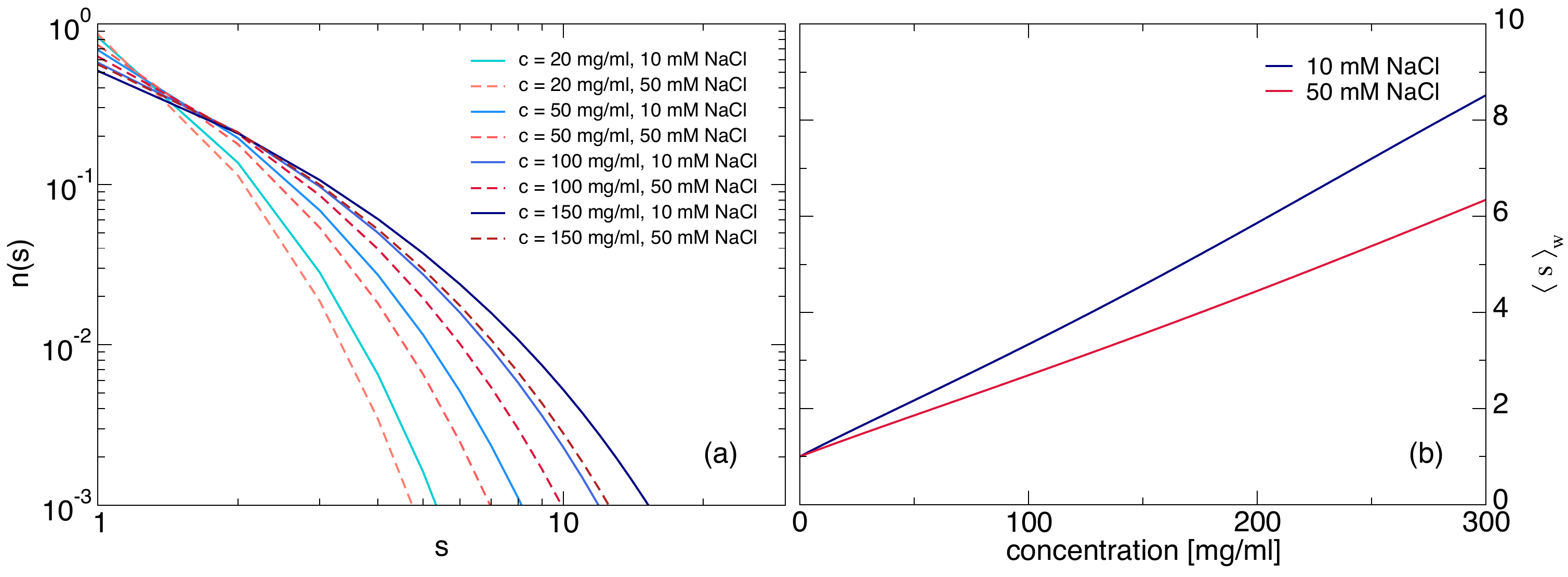}
	\caption{\footnotesize (a) Cluster size distribution $n(s)$ as a function of the cluster size $s$ for different mAb concentrations based on the parameters from a Wertheim analysis of the SLS data for 10 mM (solid lines) and 50 mM (dashed lines) NaCl. (b) Weight average aggregation number $\left\langle s \right\rangle_w$ as a function of concentration for the parameters from the Wertheim analysis for 10 mM (blue line) and 50 mM (red line) NaCl.
	}
	\label{sizedistrib}
\end{figure}

While Wertheim theory provides us with a method capable of calculating the osmotic compressibility or apparent aggregation number that can be compared with the SLS data, it does not allow us to calculate other experimental quantities such as those obtained by microrheology or SAXS. To this aim, we need the distribution $n(s)$ of clusters of size $s$ as a function of concentration $c$. We therefore use the fact that our YAB model belongs to a class of so-called hyperbranched polymers~\cite{Rubinstein2003}, which allows us to calculate the full cluster size distribution at each concentration and solvent condition using hyperbranched polymer theory (hpt) as was previously described in Ref.~\cite{Skar-Gislinge2019}.
In hpt terminology, a YAB molecule corresponds to a functionality type $AB_{f-1}$ with functionality $f=3$, for which the bond probability $p$ of Wertheim theory is the fraction of bonded $B$ groups and $(f-1)p$ the fraction of bonded $A$ groups. For hyperbranched polymers, there is one non-bonded $A$ group for each cluster, which implies that the average number of monomers per cluster is the reciprocal of the fraction of unreacted $A$ groups. Hence, the only input needed to evaluate $n(s)$ is the bond probability $p$, which we directly get from Wertheim theory. In the YAB model, calling $p$ ($2p$) the fraction of $B$ ($A$) patchy sites, the cluster size distribution $n(s)$ in the framework of hyperbranched polymer theory is finally given by
\begin{equation}\label{ns}
n (s) = \frac{(2s)!}{s!(s+1)!} p^{s-1}(1-p)^{s+1}.
\end{equation}
Therefore, $n(s)$ is the probability of finding clusters of size $s$ for a system with bond probability $p$. The corresponding cluster size distributions obtained with the parameters from the Wertheim analysis are shown in Fig. \ref{sizedistrib}(a) for four different concentrations and both ionic strengths. We see that the resulting cluster size distributions are very broad, resulting in significantly different values for different weighted averages as pointed out already earlier \cite{Skar-Gislinge2019}. This is important when considering results from different methods such as SLS, DLS or rheology, which all provide differently weighted average values. The corresponding results for the weight-average aggregation number $\big\langle s \big\rangle_w = \sum n(s) s^2/\sum n(s) s$ as a function of concentration are also shown in Fig. \ref{sizedistrib}(b). 

\subsection{Predicting SAXS data for self-assembling mAbs}

Having theoretical descriptions for the concentration dependence of both $n(s)$ and $\big\langle s \big\rangle_w$, we now make an attempt to reproduce the experimental data. The goal is to develop analytical models that allow us to calculate scattering intensities and structure factors of mAb solutions that undergo self-association into concentration-dependent transient clusters. 
The scattering intensity measured in a SAXS experiment can be written as \cite{Schurtenberger1993}
\begin{equation}
	\label{Iq}
	\frac{1}{cKM_1} \frac{d\sigma}{d\Omega}(q) = \big\langle s \big\rangle_w \big\langle P_c(q) \big\rangle_w S^{eff}_c(q),
\end{equation}

\noindent where $c$ is the weight concentration, $K$ is a contrast term that primarily reflects the excess electron density between mAb and solvent, $M_1$ is the molar mass of an individual mAb (i.e., a monomer), $d\sigma(q)/d\Omega$ is the normalized $q$-dependent scattering intensity, 
$\big\langle P_c(q) \big\rangle_w$ is the intensity-weighted average form factor of the clusters and $S^{eff}_c(q)$ is the effective structure factor of the cluster fluid.

In order to reproduce the measured scattering intensity for the mAb solutions, there are thus two tasks, namely (i) to calculate the cluster form factor $P_c(q)$ as a function of the aggregation number $s$, 
and (ii) to find an appropriate model and expression for the effective structure factor $S^{eff}_c(q)$ for the cluster fluid at the different concentrations.
\begin{figure}[b!]
	\centering
	\includegraphics[width=1\linewidth]{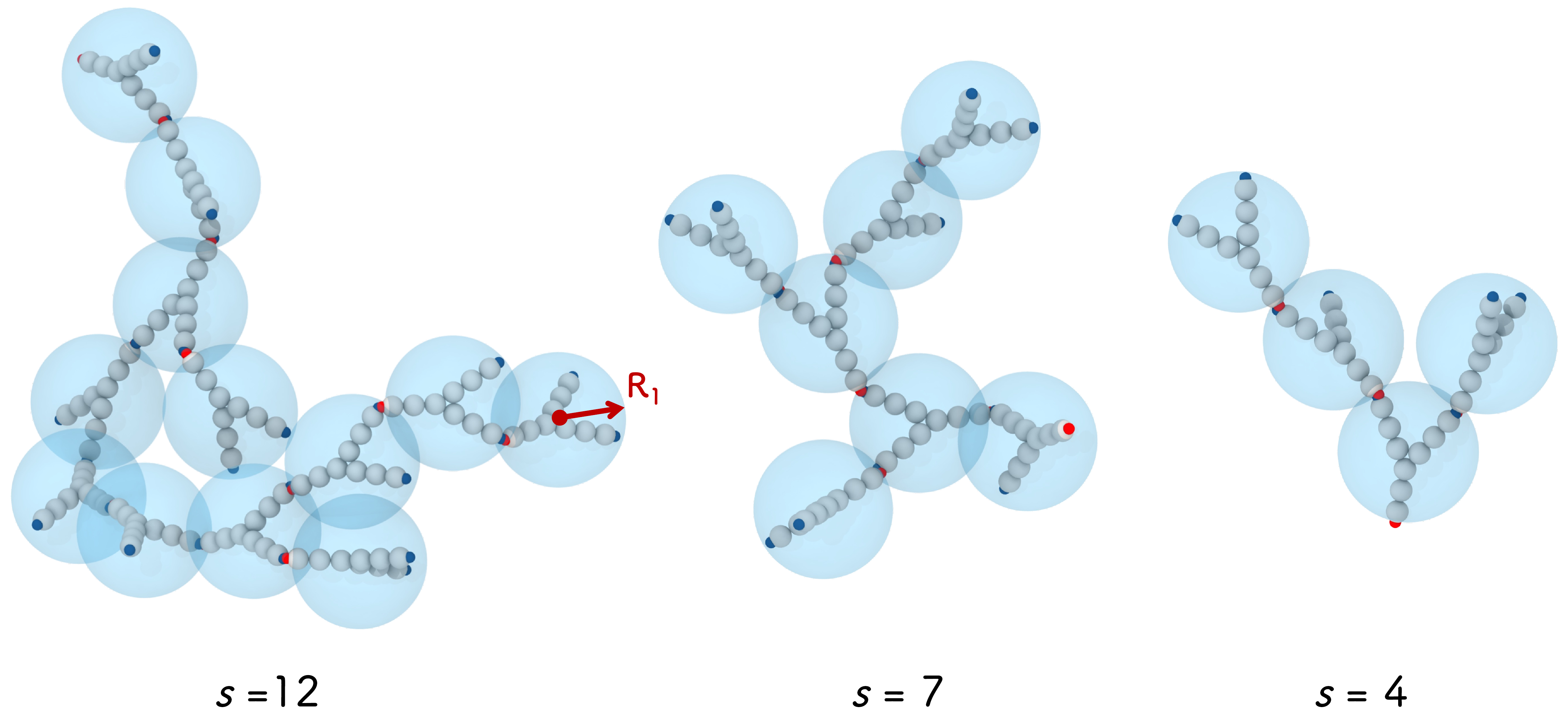}
	\caption{\footnotesize Schematic view of the coarse grained cluster model used to calculate the cluster form factor $P_c(q)$. Shown are examples of mAb clusters with $s=12, 7$ and $4$, from simulations of rigid Ys, where each mAb monomer is modelled as a rigid Y consisting of 9 spheres (see Figure~\ref{fig:model}(b)), and the further coarse grained cluster where each mAb monomer is modelled as a sphere of radius $R_1$.}
	\label{fig:cluster}
\end{figure}
Here we follow two approaches, both based on a coarse-grained view of the mAb clusters, as schematically drawn in Figure~\ref{fig:cluster}. Clusters are either modeled as assembled from patchy hard Ys formed by 9 spheres 
(see Fig. \ref{fig:model} and Methods) or then, after a further coarse-graining step, as made up of spheres with radius $R_1$, where the subscript $1$ indicates that this is the radius of a monomer ($s=1$, see Figure~\ref{fig:cluster}).
The average form factor of single clusters formed by $s$ monomers can then be directly calculated using

\begin{equation}
	\label{Debye}
	P_c(q)= \frac{1}{s^2} P_1(q) \sum_i \sum_j \left\langle  \frac{sin(qr_{ij})}{qr_{ij}}\right\rangle,
\end{equation}

\noindent where $P_1(q)$ is the form factor of the monomer ($s=1$), $r_{ij}$ is the center-center distance between monomer $i$ and $j$, and $\left\langle \cdots \right\rangle$ denotes an average over clusters with different conformations. For the calculation of $P_c(q)$ we rely on different forms for $P_1(q)$: (i) for a direct comparison with experimental data we use the measured form factor of the mAb; (ii) we consider a cluster of Ys or spheres model as shown in Figure~\ref{fig:cluster}  and use the form factor of a 9-bead Y or of a sphere with radius $R_1$. Thus, we can rewrite the term appearing on the right-hand side of Eq.~\ref{Debye} as,
\begin{equation}
    \frac{1}{s^2}  \sum_i \sum_j \left\langle  \frac{sin(qr_{ij})}{qr_{ij}}\right\rangle= \frac{1}{s} S_c(q),
\end{equation} 
where $S_c(q)$ is the structure factor of the cluster with a normalization that yields $S_c(0) = s$ and $S_c(\infty) = 1$. Using this notation, the scattering intensity $I_c(q)$ from a single cluster can be written as 
\begin{equation}
    I_c(q) = s P_1(q) S_c(q) = s^2 P_c(q).
\end{equation}

In the next step, we thus develop a model that provides us with an explicit description of $S_c(q)$. This can either come from \textit{computer simulations} or from \textit{analytical models} that yield $P_c(q)$ (or $S_c(q)$) as a function of $s$.

\subsubsection{Using \textit{computer simulations} of a colloid-inspired antibody model}

In order to obtain a microscopic description of the antibody assembly and thus the structure factor for single clusters, we run Monte Carlo simulations of an ensemble of antibodies, described by the 9-bead model depicted in Fig.~\ref{fig:model}(b). To allow the antibodies to self-assemble, the temperature is fixed to $T=0.11$, as determined from Wertheim theory. More details on such simulations and on the conversion between real and simulation units are provided in Methods.

\begin{figure}[t!]
	\centering
	\includegraphics[width=1\linewidth]{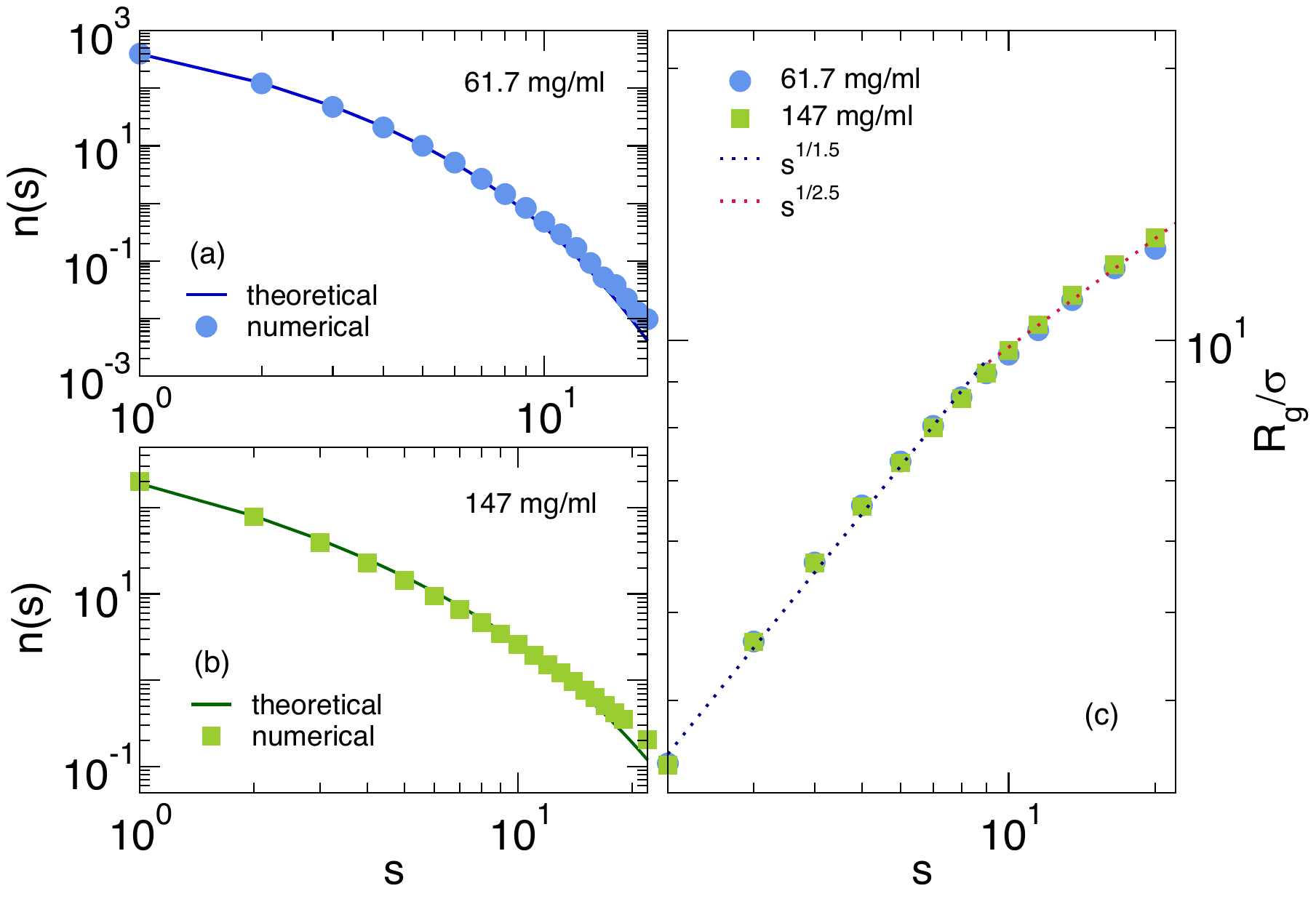}
	\caption{\footnotesize (a, b) Cluster size distribution $n(s)$ as a function of the cluster size $s$ calculated from simulations of the 9-bead patchy model for $c=61.7$ mg/ml and for $c=147$ mg/ml, respectively, compared to the corresponding hpt predictions;  (c) average radius of gyration $R_g$ for clusters of different sizes $s$ for the same mAb concentrations as in (a, b). The two dotted lines are fits to the small and large cluster sizes with $R_g \sim s^{1/d_F}$: for $s \lesssim 10$, $R_g \sim s^{1/1.5}$, while for larger sizes $R_g \sim s^{1/2.5}$.}
	\label{fig:clustersimulations}
\end{figure}

We first study the cluster size distribution $n(s)$ as a function of the cluster size $s$. Antibodies belong to the same cluster when a patch of type A on the first antibody and a patch of type B of the second one are linked, that is when A-B patches are closer than the square-well attraction distance $r=0.2633\sigma$. This is reported in Fig.~\ref{fig:clustersimulations}(a, b) for two concentrations $c=61.7$ and $147$ mg/ml, respectively. Together, we also plot the corresponding theoretical predictions from hpt. As expected, we find that the clusters are rather polydisperse in size, reaching $s>20$ at the highest concentration. Representative simulation snapshots are shown in Fig.~\ref{fig:cluster} for $s=12, 7$ and $4$. We note that larger clusters are also found  but they are beyond our numerical accuracy, since their number is $< 0.1\%$ of the total. The agreement between numerical data and theoretical predictions is overall good for both concentrations, thus confirming that the employed model does follow hyperbranched polymer theory.  Small deviations at large $s$ are due to the minimal presence of multiple bonds for the same patch, which are not taken into account in the theoretical treatment. However, their contribution is negligible for both concentrations, implying that we can use the theoretical prediction to evaluate $n(s)$ at all concentrations. 

It is interesting to compare the cluster size distributions obtained here with those presented in earlier studies of the link between antibody solution viscosity and self assembly using different coarse grained computer simulations with alternative bead models~\cite{Chowdhury2020, lai2021calculation}. Our patchy model is not expected to undergo percolation, where antibodies form a single system-spanning cluster, at any finite concentration. This is due to the fact that the model used belongs to the hyperbranched class of polymers that results in branched structures without gelation~\cite{rubinstein2003polymer}. We have therefore concluded that the strong increase of the relative viscosity due to mAb self-assembly is thus not related to a sol-gel transition at a sufficiently high concentration, but attributed it to a cluster glass transition as found for hard or attractive hard sphere colloids~\cite{Skar-Gislinge2019}. We do not expect this to be generic for all antibodies, and it is worth pointing out that computer simulations using a different bead model indeed resulted in the formation of a percolating large cluster at high concentrations~\cite{lai2021calculation}.

From simulations, we can then assess the shape of single clusters. To this aim, we calculate the average radius of gyration $R_g$ for clusters of the same size $s$, that is reported in Fig.~\ref{fig:clustersimulations}(c) as a function of size for the same concentrations as in panels a and b. From this plot, we can extract information on the clusters fractal dimension $d_F$, since $R_g \sim s^{1/d_F}$. We identify two different regimes, for clusters smaller and bigger than $s \approx 10$:  in the first range, we find $d_F \approx 1.5$, while for larger clusters a fractal dimension $d_F \approx 2.5$ is compatible with the data.

\begin{figure}[t!]
	\centering
	\includegraphics[width=0.6\linewidth]{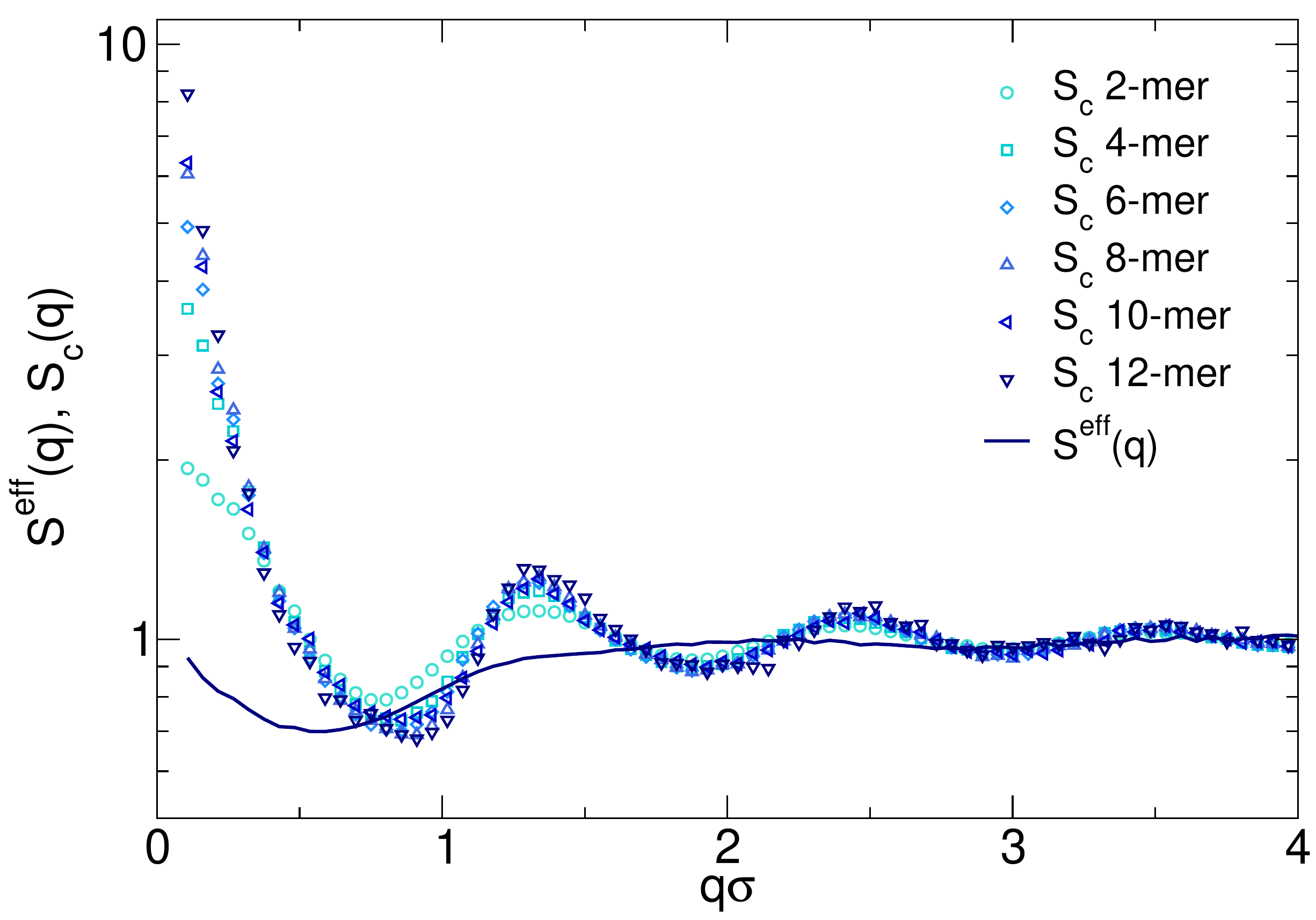}
	\caption{\footnotesize Cluster structure factors $S_c(q)$ (Eq.~\ref{Sqc-sim}) for $s = 2, 4, 6, 8, 10, 12$ and total structure factor $S^{eff}_{sim}(q)$ (Eq.~\ref{Sqtot-sim}) obtained from computer simulations for $c = 61.7$ mg/ml. Data are shown in simulation units (where $\sigma$ is the bead size).
	}
	\label{fig:scq-onlysim}
\end{figure}
For each cluster size $s$, we also calculate the corresponding cluster structure factors $S_c(q)$ as
\begin{equation}
\label{Sqc-sim}
 S_c(q) = \frac {1}{s}\left\langle 
 \sum_{i,j=1,s} e^{-i {\bf q} \cdot ({\bf r}_{i,cm}-{\bf r}_{j,cm})} \right\rangle,
\end{equation}
where  ${\bf r}_{i,cm}$ and ${\bf r}_{j,cm}$ are the coordinates of the centers of mass of $i$-th and $j$-th Y-molecule within the same cluster and the average is taken over all clusters of size $s$ in the whole simulation trajectory.
The structure factors for clusters of different sizes are reported in Fig.~\ref{fig:scq-onlysim} for $c = 61.7$ mg/ml. With increasing $s$, the first peak of $S_c(q)$ becomes more pronounced, accompanied by an increase of its signal at low wavenumbers. In addition, simulations allow us also to calculate the total effective structure factor of the system $S^{eff}_{sim}(q)$. This is the analogue of the experimentally measured $S^{eff}(q)$ and it is defined as,
 \begin{equation}\label{Sqtot-sim}
\begin{split}
S^{eff}_{sim}(q) &= \frac{1}{P_1(q)} \frac{1}{9N}\left\langle \sum_{i_b=1}^{9N}\sum_{j_b=1}^{9N} e^{-i {\bf q} \cdot ({\bf r}_{i_b}-{\bf r}_{j_b})} \right\rangle \\
\end{split}
\end{equation}
where now the sum is taken over all beads of all antibodies, whose coordinates are $\textbf{r}_{i_b}, \textbf{r}_{j_b}$, including cross-interactions and the average is taken over all trajectories. In addition, $P_1(q)$ is calculated from simulations of a single Y.

The total structure factor calculated in this way is also shown in Fig.~\ref{fig:scq-onlysim} and, strikingly, it shows very little oscillations and almost no peaks, except for a slight increase at small $q$. These calculations will be compared in later sections with analytical calculations and experimental results to provide a comprehensive description of the solution structure.

\subsubsection{Using \textit{polymer theory} to calculate the cluster form factor}

To develop an analytical model for $S_c(q)$, and given the relatively open structure of the clusters found in simulations, we first use a simple polymer model, where we assume that the conformational average of the internal distances $r_{ij}$ is given by a freely jointed chain (fjc) model \cite{doi1988,Rubinstein2003,  Schweins2004}.
In this model for the conformation of a polymer chain of size $s$, we assume that the chain consists of $s$ monomers linked by $s-1$ bonds of length $b$ that are able to point in any direction independently of each other, i.e. with no correlation between the direction of different bonds. The average radius of gyration of such a chain is thus given by a scaling law of the form $\left\langle R_g \right\rangle \sim s^{1/2}$ \cite{doi1988, Rubinstein2003}. The conformations described by the fjc model would thus be compatible with a fractal cluster structure with $d_F = 2.0$, that is intermediate between the fractal dimensions found in our simulations for small and large cluster sizes.
This implies that in Eq. \ref{Debye}
\begin{equation}
	\label{fjc-model}
	  \left\langle  \frac{sin(qr_{ij})}{qr_{ij}}\right\rangle = \left( \frac{sin(qb)}{qb} \right) ^{\vert j-1\vert}
\end{equation}
\noindent where b is the distance between two spheres, i.e. the sphere diameter $2R_1$ in our model. Evaluating the double sum finally results in the following expression for the cluster form factor in the fjc approximation:

\begin{equation}
	\label{fjc-model2}
	 P_{c,fjc}(q)= \frac{2 P_1(q)}{s^2}   \left[ \frac{s}{1 - sin(qb)/qb} - \frac{s}{2} - \frac{1 - (sin(qb)/qb)^{s}}{(1 - sin(qb)/qb)^2} \frac{sin(qb)}{qb} \right]
\end{equation}

\begin{figure}[b!]
	\centering
	\includegraphics[width=1\linewidth]{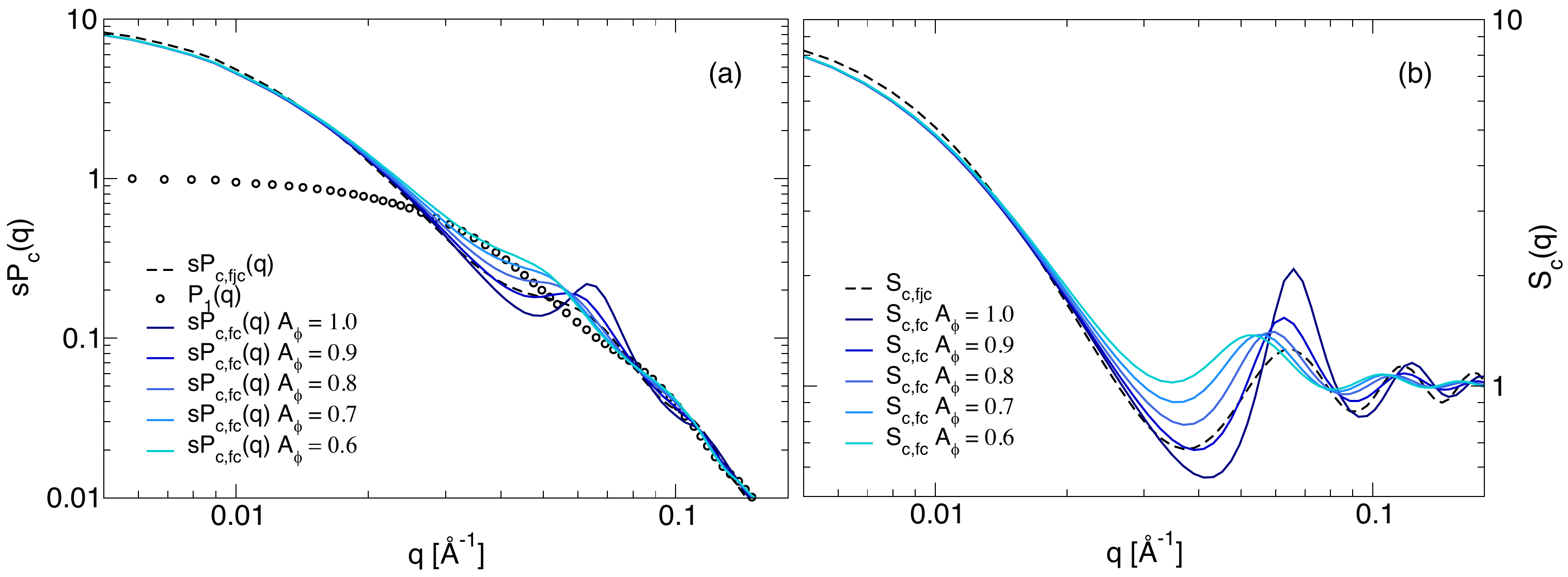}
	\caption{\footnotesize (a) Normalized intensity $s P_c(q)$ for a mAb cluster with aggregation number $s = 10$ using the freely jointed chain model (Eq. \ref{fjc-model2} and $b = 12$ nm, black dashed line), and a fractal cluster model (Eqs. \ref{Scluster} and \ref{Fisher-Burford} and a hard sphere structure factor with $R_1 = 6$ nm) for different values of the internal volume fraction ($A_{\phi} =$ 1.0, 0.9, 0.8, 0.7 and 0.6 (blue lines)), and the experimentally measured mAb form factor $P_1(q)$ obtained at a concentration of 4.9 mg/ml (circles). (b) The corresponding cluster structure factors $S_c(q)$ for the same models.
	}
	\label{fig:FJC-model-Pc}
\end{figure}

An example of the resulting cluster intensity $s P_{c,fjc}(q)$ using Eq. \ref{fjc-model2} is shown in Fig. \ref{fig:FJC-model-Pc}(a) (dashed line) for a mAb cluster with $s = 10$ and $b = 12$ nm. This value of $b$ was chosen based on the actual geometrical dimensions of the mAb and corresponds approximately to the diagonal distance between the positively and negatively charged ends, i.e. to the expected bond length in our model. The chosen normalization allows us to directly compare the cluster form factor $s P_{c,fjc}(q)$ with the measured monomer form factor $P_1(q)$. We see that at low $q$-values the overall scattering pattern is dominated by the overall cluster size. At higher $q$-values, $P_{c,fjc}(q)$ approaches the monomer form factor, modulated however with the local correlations between individual monomer beads in the cluster expressed by the cluster structure factor in the fjc approximation $S_{c,fjc}(q)$, shown in Fig.~\ref{fig:FJC-model-Pc}(b).

\subsubsection{Using \textit{colloid theory} to calculate the cluster form factor}

We also develop a second coarse-grained model for $S_c(q)$ that is instead based on colloid theory \cite{Larsen2020}. Here we start from the cluster structure factor of a single fractal colloid cluster (fc) $S_{c,fc}(q)$ of size $s$ given by the double sum in Eq.~\ref{Debye}, which we rewrite according to

\begin{equation}
	\label{Debye2}
	S_{c,fc}(q)= \frac{1}{s} \sum_i \sum_j \left\langle  \frac{sin(qr_{ij})}{qr_{ij}}\right\rangle = 1 + \frac{1}{s} \sum_i \sum_{j \neq i} \left\langle  \frac{sin(qr_{ij})}{qr_{ij}}\right\rangle
\end{equation}

\noindent where the last term in Eq. \ref{Debye2} is the cross term between the individual monomers in the cluster and the large embedding sphere with radius $R_c$. This term can be rewritten as

\begin{equation}
	\label{Debye3}
	S_{c,fc}(q)= 1 + (s - 1) P_L(q)
\end{equation}

\noindent where $P_L(q)$ is the form factor of the embedding sphere (or cluster). Eq. \ref{Debye3} does not take into account correlations between monomers within the cluster, which can for example be considered by introducing a hard sphere structure factor $S_{HS}(q,\phi_{int})$, where the internal volume fraction of a cluster of radius $R_c$ is given by $\phi_{int} = s (R_1/R_c)^3$

\begin{equation}
	\label{Scluster}
	S_{c,fc}(q)= S_{HS}(q,\phi_{int}) + (s - S_{HS}(q,\phi_{int})) P_L(q)
\end{equation}

In a final step, we need to select appropriate models for $S_{HS}(q,\phi_{int})$ and $P_L(q)$. For $S_{HS}(q,\phi_{int})$ we can for example use liquid state theory and the corresponding structure factor for hard spheres given by the Percus-Yevick (PY) expression \cite{Hansennew}. For $P_L(q)$ we choose the Fisher-Burford expression that has been used to describe the scattering intensity of fractal clusters with fractal dimension $d_F$, \cite{Fisher1967, Sorensen1999}

\begin{equation}
	\label{Fisher-Burford}
	P_L(q)=  \left(1 +  \frac{2 R_g^2 q^2}{3d_F} \right)^{-d_F/2} .
\end{equation}

\noindent In order to calculate the cluster structure and form factors for this model, we also need to determine the radius of gyration $R_g$ and the internal volume fraction $\phi_{int}$. To be internally consistent, we have used the common relationship for the radius of gyration of a fractal cluster $R_g^c$ given by

\begin{equation}
	\label{Rg-cluster}
	R_g^c = R_1 \left(\frac{s}{k}\right)^{1/d_F}
\end{equation}

\noindent where $R_1$ is the monomer size and $k$ a constant that depends on the fractal dimension $d_F$. The internal volume fraction $\phi_{int}$ is thus given by

\begin{equation}
	\label{phi-internal}
		\phi_{int} = A_{\phi} s \left(\frac{R_1}{R_g^c}\right)^{3}
\end{equation}

\noindent where the parameter $A_{\phi}$ corrects for the fact that the monomers in the fractal cluster are treated as spheres with size $R_1$, whereas their effective hard sphere radius and thus their excluded volume is smaller than $R_1$ due to the Y-shape of the mAb.

We again use $s = 10$ and choose $d_F = 2.5$ in agreement with the computer simulation results for larger clusters, which in turn yields $k = 0.71$. The resulting cluster structure factor does depend on the value of $R_1$, as this determines both the low-$q$ behavior through the overall cluster size $R_g^c$ as well as the position of the nearest neighbor correlation peak roughly given by $q^* \approx 2 \pi/2 R_1$ (note that the internal volume fraction is independent of the choice of $R_1$, since it only depends on the ratio $R_1/R_c$). Moreover, the internal correlation peak will also depend on the internal volume fraction due to the concentration dependence of the structure factor of hard spheres calculated for example via the Percus-Yevick (PY) expression \cite{Hansennew}. 
 
The results for both $s P_{c,fc}(q)$ and $S_{c,fc}(q)$ are also shown in Fig.~\ref{fig:FJC-model-Pc} for different values of $A_{\phi}$, together with the results of the fjc model.
Due to the similar $R_g^c$,  the low-$q$ part almost overlap for both models. In fact, $R_g^c = 2 R_1 (s/6)^{1/2} = 15.5$ nm for the fjc model \cite{Rubinstein2003} and $R_g^c = 17.4$ nm for the fc model.
Since both models have a slightly different asymptotic slope given by $1/d_F$, with $d_F = 2$ for the fjc and $d_F = 2.5$ for the fc model, this then results in a very similar initial $q$-dependence that would be difficult to distinguish in real experimental data. However, at higher $q$-values, differences become much larger. Nearest neighbor correlations for the fc model are strongly dependent on the internal volume fraction, which becomes highlighted when looking at $S_{c,fc}(q)$ for different values of $A_{\phi}$. For the fjc model, longer range correlations persist due to the underlying linear chain structure with a well defined bond length, while these decay more quickly for the fc model. 

\subsubsection{Comparison between fjc and fc models, and computer simulations}

\begin{figure}[t!]

	\includegraphics[width=1\linewidth]{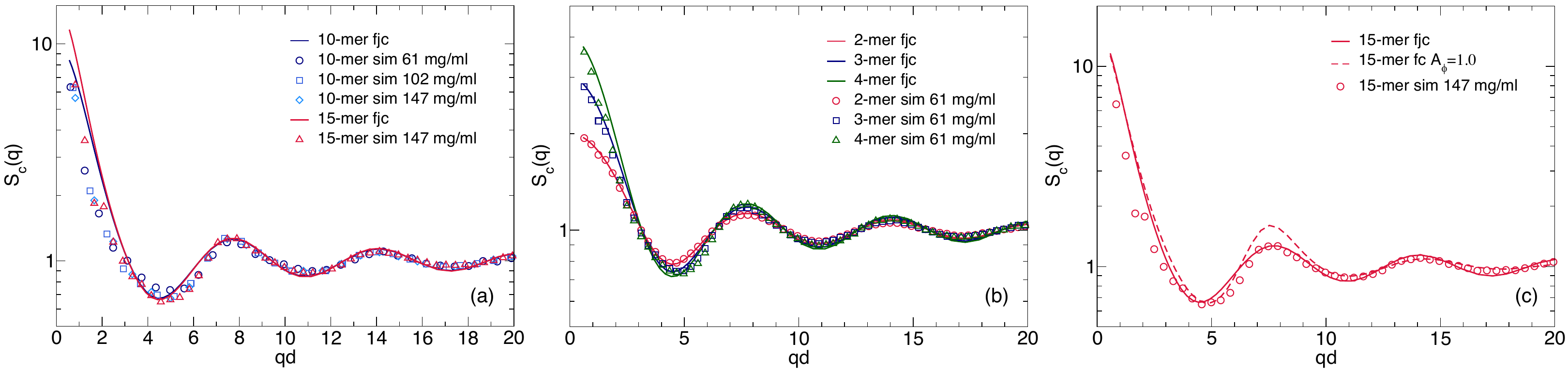}
	\caption{\footnotesize (a) Comparison of the normalized cluster structure factors $S_c(q)$ for $s = 10$ obtained from computer simulations for a hard 9 bead Y model at three different concentrations corresponding to $c = 61.7$, 102.2 and 147.3 mg/ml (blue symbols), respectively. Also shown are the results for $s = 15$ from simulation at $c = 147.3$ mg/ml (red triangles), and from the fjc model for $s = 10$ (blue line) and $s = 15$ (red line). (b) Comparison of the cluster structure factors $S_c(q)$ obtained from computer simulations for a hard 9 bead Y model at $c = 61.7$ mg/ml and $s = 2$ (red circles), $s = 3$ (blue squares) and $s = 4$ (green triangles), together with those calculated with the fjc model for the same cluster sizes ($s = 2$ (red line), $s = 3$ (blue line), $s = 4$ (green line)), respectively. (c) Comparison of the cluster structure factors $S_c(q)$ for $s = 15$ obtained from computer simulations for a hard 9 bead Y model at $c = 147.3$ mg/ml (red circles), the fjc (red solid line) and the fc models (Eqs. \ref{Scluster} and \ref{Fisher-Burford} and a hard sphere structure factor with $R_1 = 6$ nm for $A_{\phi} =$ 1.0) (red dashed line), respectively.
	}
	\label{fig:FJC-9beadMC-Sq}
\end{figure}

We can now compare the cluster structure factors $S_c(q)$ obtained by the two models with those calculated from the computer simulations using Eq.~\ref{Sqc-sim}. To this aim, the individual $S_c(q)$ are reported as a function of $qd$, i.e. normalized by the effective distance $d$ between different mAbs in the clusters given either by the bead size $d = b = 12 nm$ or the diagonal distance between the oppositely charged patches given by $d = 5.8 \sigma$, respectively. In order to test the absence of concentration effects on $S_c(q)$, we compare the data with the results from simulations at three different concentrations corresponding to $c = 61.7$ mg/ml, 102.2 mg/ml and 147.3 mg/ml, respectively. As shown in Fig.~\ref{fig:FJC-9beadMC-Sq}(a),
the three different $S_c(q)$ obtained for $s = 10$ overlap within the statistical errors, indicating that the average structure of the clusters formed are independent of concentration for a given value of the cluster size $s$.
The agreement between the simulation results and those obtained from the fjc model are also very good. 
Fig. \ref{fig:FJC-9beadMC-Sq}(a) furthermore illustrates that for sufficiently large cluster sizes the internal structure described by $S_c(q)$ becomes independent on $s$ except for low $q$-values, where $S_c(q)$ approaches $s$. 

However, for small cluster sizes, the internal structure starts to strongly depend on $s$ as illustrated in Fig. \ref{fig:FJC-9beadMC-Sq}(b) for $s = 2$, 3 and 4, respectively. Here we also see that for these small cluster sizes the results from simulations and the fjc model completely overlap and the two approaches are now identical within error bars.

Finally, we show in Fig. \ref{fig:FJC-9beadMC-Sq}(c) the cluster structure factors for both fjc and fc models and simulations for $s = 15$. While the model based on fractal colloidal clusters is obviously not suitable for small cluster sizes since a fractal description is not adequate, it does however agree quite well with the computer simulations and the fjc model at sufficiently large values of $s$.

\subsubsection{Solution structure factor}

In order to calculate the total normalized scattering intensity for a cluster fluid as described by Eq. \ref{Iq}, we finally also need a model for the effective structure factor $S^{eff}_c(q)$ of the cluster fluid. Given the very broad size distribution of the self-assembled antibody clusters at higher concentrations as predicted by either hyperbranched polymer theory (hpt) or our coarse grained simulations, we do expect very weak structural correlations even at the nearest neighbor distance, similar to what would be found for example for polymer solutions. We therefore use a so-called random phase approximation (RPA), where the structure factor is given by \cite{Higgins1996}

\begin{equation}
	\label{RPA}
	S^{eff}_c(q) = \frac{1}{1 + \frac{1 - S^{eff}_c(0)}{S^{eff}_c(0)}\left\langle P_c(q) \right\rangle_w}. 
\end{equation}

We note that $S^{eff}_c(0) = S^{eff}(0)/\langle s\rangle_w$ now corresponds to the effective structure factor of a solution of polydisperse spheres, reflecting the fact that the mAb clusters (and not the individual antibodies) are the new objects of interest. In this way, the total normalized scattering intensity given by Eq. \ref{Iq} can then be rewritten as

\begin{equation}
	\label{Iq-th}
	\frac{1}{cKM_1} \frac{d\sigma}{d\Omega}(q) = \frac{ \langle s\rangle_w \left\langle P_c(q) \right\rangle_w}{1 + \frac{1 - S^{eff}_c(0)}{S^{eff}_c(0)} \left\langle P_c(q) \right\rangle_w}. 
\end{equation}
In a next step, we need to calculate $\langle s\rangle_w$ and $S^{eff}_c(0)$ as a function of concentration based on our previously established approach using a combination of Wertheim theory and hpt. 
Here, we use the obtained bond probability versus concentration relationship and perform a next coarse graining procedure where we treat the clusters as the new hard or sticky colloids. We thus first make use of the cluster size distributions and the weight average aggregation number $\langle s\rangle _w$ previously calculated at all concentrations with hpt (see Fig. \ref{sizedistrib}). Assuming hard or sticky hard sphere-like interactions between the different clusters, we can then calculate the concentration dependence of the apparent weight average aggregation number $\langle N_{app} \rangle_w $, given by 

 \begin{equation}
 \label{naggapp_col}
	\langle N_{app} \rangle_w  = \langle s \rangle_w S^{eff}_c(0).
	\end{equation}

We use the same conversion of the weight concentration into number densities of mAbs in units $\sigma^{-3}$ based on $\sigma = 2.72$ nm, and then calculate the number densities of clusters using $\rho_{cluster} = \rho/\langle s\rangle_n$, where $\langle s\rangle_n = \sum n(s) s/\sum n(s)$ is the number-average aggregation number.
In doing these calculations we also have to reconsider the effective hard sphere cluster volume fraction $\phi_{HS}$. Starting point for calculating $\phi_{HS}$ is the hard sphere volume fraction used in the Wertheim analysis $\phi = \rho V_{hs}$. We then allow for an additional scaling parameter $A$ 
and also take into account that clusters are fractal, giving \cite{Skar-Gislinge2019}
\begin{equation}\label{phiHS}
\phi_{HS} = A \phi \ \langle s\rangle_n^{(3 - d_F)/d_F},
\end{equation} 
where $d_F$ is the fractal dimension of the clusters and $\phi$ is the nominal antibody volume fraction used in the Wertheim analysis. Given the small cluster sizes with $\langle s\rangle_n < \langle s\rangle_w < 10$ for all concentrations investigated (see Fig. \ref{sizedistrib}(b)), we use $d_F = 2.0$.

In our earlier investigations at low ionic strength, we found best agreement for a model of clusters that interact as sticky hard spheres, for which the low scattering vector limit of the effective static structure factor $S^{eff}_c(0)$ becomes~\cite{Skar-Gislinge2019}

\begin{equation}
	\label{S(0)-stsph}
	S_{SHS}(0) = \frac{(1 - \phi_{HS})^4}{(1 + 2 \phi_{HS} - \lambda \phi_{HS})^2},
\end{equation}

\noindent
with
\begin{equation}\label{lambda}
	\lambda = 6 (1 - \tau + \tau / \phi) \biggl(1-\sqrt{1 - \frac{1 + 2 / \phi}{6 (1 - \tau + \tau / \phi)^2}}\biggr).
\end{equation}

\noindent where $\tau$ is the stickiness parameter that is inversely proportional to the strength of the attractive interaction. \cite{Piazza1998,Cichocki1990} Together with the concentration dependence of $s$, obtained with Wertheim theory and hpt, we can then calculate $\langle N_{app} \rangle_w $ using Eq.~\ref {naggapp_col} for both ionic strengths values investigated.

The corresponding best fit results using $A = 1.4$ and $\tau = 2.5$ are shown in Figure \ref{Fig1-Napp-etar}.  
The agreement with experimental data is indeed excellent, assuming that in the coarse grained model we have an effective hard sphere volume fraction that is $\approx40$\% higher than for the individual mAbs in the Wertheim analysis. Given that clusters cannot overlap as much as individual antibodies do, this estimate does appear to be realistic.

A further consistency check of this additional coarse graining step can also be obtained from the microrheology data. Here we calculate the concentration dependence of the relative viscosity $\eta_r$. Theoretical work on hard sphere and attractive systems using mode coupling theory (MCT) and computer simulations predicts a power-law dependency of the reduced viscosity 

\begin{equation}\label{powerlaw}
\eta_r = \left(\frac{\phi_g - \phi_{HS}}{\phi_{g}}\right)^{-\gamma}
\end{equation}

\noindent in the vicinity of the arrest transition, where $\phi_g$ is the maximum packing fraction, which depends on the polydispersity of the system and the strength of the attraction.\cite{Puertas2007} The value of $\gamma$ depends on the interaction between particles, with typical values being $\gamma \sim 2.8$ for hard spheres and $\gamma \geq 3$ for attractive particles.\cite{Puertas2007, Puertas2003} The viscosity data obtained for the mAb solutions at both ionic strengths are well reproduced with a power law fit with two fit parameters, namely an exponent $\gamma = 3.0$ and the arrest packing fraction $\phi_\textrm{g} = 0.63$ (see Fig. \ref{Fig1-Napp-etar}(b)). The latter value is consistent with expectations that arrest of the mAb clusters is dominated by excluded-volume interactions, providing further support for the calculated dependence of the clusters size on concentration. 
Therefore, our simple model is capable of predicting quantitatively the measured $c$ and ionic strength dependence based on SLS experiments. In this context, it is also interesting to compare the calculations for the case of self-assembling antibodies with an estimate of $\eta_r$ for a hypothetical case where the mAbs do not assemble into clusters and where $\eta_r$ is given by Eq. \ref{powerlaw}, but using the hard sphere volume fraction from the Wertheim analysis instead. The resulting values are also shown in Fig. \ref{Fig1-Napp-etar}(b) for two values of the exponent $\gamma$ (2.8 and 2.0, the latter corresponding to the often used Quemada relation for hard spheres \cite{Quemada1977}) and demonstrate the dramatic effect of cluster formation at higher mAb concentrations.

We then use the results from this analysis to calculate the full $q$-dependence of the total normalized intensity of the cluster fluid using Eq. \ref{Iq-th}. However, instead of plotting the intensity, we calculate an effective measured solution structure factor $S^{eff}(q)$ where we divide the total normalized intensity with the monomer form factor, i.e.

\begin{equation}
	\label{Sq-th}
	S^{eff}(q) = \frac{1}{cKM_1} \frac{d\sigma}{d\Omega}(q) \frac{1}{P_1(q)}= \frac{\left\langle S_c(q)\right\rangle_w}{1 + \frac{1 - S^{eff}_c(0)}{S^{eff}_c(0)}\left\langle P_c(q) \right\rangle_w} =   \left\langle S_c(q)\right\rangle_w S^{eff}_c(q).
\end{equation}

\noindent where $\left\langle S_c(q)\right\rangle_w$ is the weight average internal cluster structure factor. We thus compare the measured $S^{eff}(q)$ with the calculated quantity $\left\langle S_c(q)\right\rangle_w S^{eff}_c(q)$. Here $\left\langle S_c(q)\right\rangle_w$ is calculated using the individual $S_c(q)$ for all cluster sizes $s$ obtained either with the fjc model (Eq. \ref{fjc-model2}) or the fc model (Eq. \ref{Scluster} and \ref{Fisher-Burford}), and then performing the calculation of the corresponding weight average using the full cluster size distribution for each concentration, and $S^{eff}_c(q)$ is obtained using the RPA model (Eq. \ref{RPA}).

\begin{figure}[t!]
	\centering
	\includegraphics[width=1\linewidth]{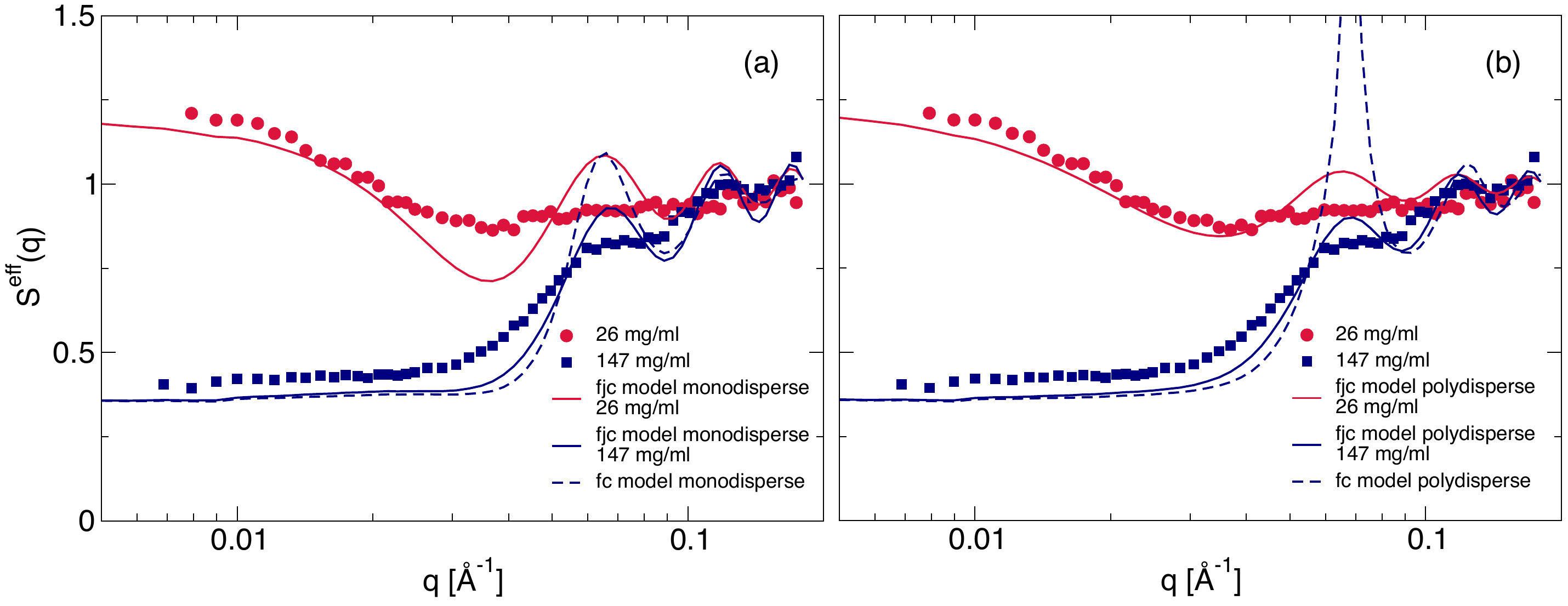}
	\caption{\footnotesize  (a) Measured effective structure factor $S^{eff}(q)$ compared with the theoretical one, calculated following Eq.~\ref{Sq-th}, where $S_c(q)$ is taken as the one corresponding to the average cluster size and $S_c^{eff}(q)$ is obtained using either the fjc  (solid lines) or fc (dashed line) models for the lowest and highest mAb concentrations measured (red: 26 mg/ml, blue: 147 mg/ml ). (b) Measured effective structure factor $S^{eff}(q)$ compared with the calculated  one, where now Eq.~\ref{Sq-th} is generalized for polydisperse systems using Eq. \ref{Iq-poly}, either for the fjc  (solid lines) or for the fc (dashed line) model for the lowest and highest concentrations measured (red: 26 mg/ml, blue: 147 mg/ml). Data are for 10 mM added NaCl. 
	}
	\label{fig:Sq-meas1}
\end{figure}

In order to demonstrate the importance of averaging all quantities over the full cluster size distribution, we perform the comparison between experimental and theoretical data in two steps. First, we calculate the effective solution structure factor based on Eq. \ref{Sq-th} using expressions for a monodisperse cluster fluid with an effective cluster size given by $\langle s\rangle _w$. The corresponding weight averaged quantities $\left\langle S_c(q)\right\rangle_w$ and $\left\langle P_c(q)\right\rangle_w$ are thus replaced by the monodisperse expressions $S_c(q)$ and $P_c(q)$,  respectively. The results of this first attempt are shown in Fig. \ref{fig:Sq-meas1}(a) for two concentrations of 26 mg/ml and 147 mg/ml, respectively, at the lower ionic strength. For these samples, the combination of Wertheim theory and hpt predicts weight average aggregation numbers of $\langle s\rangle _w = 1.62$ for 26 mg/ml and $\langle s\rangle _w = 4.5$ for 147 mg/ml, respectively. Using the sticky hard sphere cluster model, this then results in values of $S_{SHS}(0) = 0.72$ for 26 mg/ml and $S_{SHS}(0) = 0.08$ for 147 mg/ml, respectively. For an aggregation number of 2 the fc model is obviously meaningless, and therefore we have only used the fjc model for the lowest concentration. While overall the initial low-$q$ part is reasonably well reproduced, the chosen models clearly overestimate the nearest neighbor correlations at higher $q$. 

Part of the large discrepancies between measured and calculated structure factors are obviously due to the fact that polydispersity has been neglected, except for the calculation of the average cluster size $\langle s\rangle _w$. Indeed, in the first step of our comparison, we obtained $S_c(q)$ using a monodisperse fjc or fc model, where the cluster size corresponds to the nearest discrete value of the theoretical weight average aggregation number. 
Since for smaller cluster sizes the internal structure depends on the aggregation number (see Fig. \ref{fig:FJC-9beadMC-Sq}), as a second step we perform new calculations based on the full cluster size distributions from hpt, starting from the theoretical normalized scattered intensity of a polydisperse cluster fluid in the absence of interactions ($S_c^{eff}(q)=1$) given by \cite{Schurtenberger1993} 

\begin{equation}
	\label{Iq-poly}
	\frac{1}{cKM_1} \frac{d\sigma}{d\Omega}(q) =   \frac{\sum_s n(s) s^2 P_c(s,q)}{\sum_s n(s) s}
\end{equation}

\noindent where $P_c(s,q)$ is the cluster form factor of a cluster with aggregation number $s$, and $n(s)$ is the normalized cluster size distribution. In a next step we then again calculate $S^{eff}(q)$ using Eq. \ref{Sq-th} for both models. The corresponding results are also shown in Fig. \ref{fig:Sq-meas1}(b).

\begin{figure}[b!]
	\centering
	\includegraphics[width=1\linewidth]{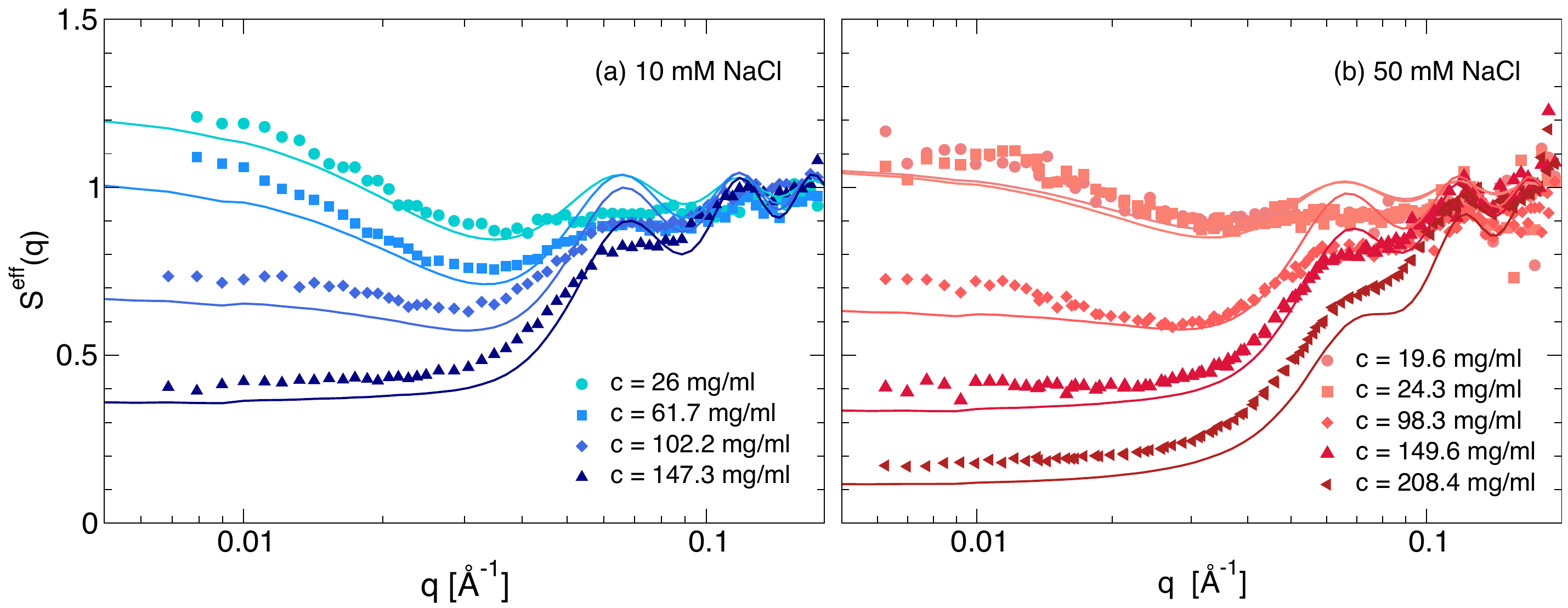}
	\caption{\footnotesize Measured effective structure factor $S^{eff}(q)$ (symbols) compared with the calculated ones based on Eqns.~(\ref{Sq-th},\ref{Iq-poly})  using the fjc model (solid lines) for different mAb concentrations measured. (a) 10 mM NaCl (blue lines: 26, 61.7, 102.2, 147.3 mg/ml); (b) 50 mM NaCl (red lines: 19.6, 24.3, 98.3, 149.6, 208.4 mg/ml).
	}
	\label{fig:Sq-meas-all}
\end{figure}

The agreement between the experimental data and the calculations for the fjc model is now improved, although the internal structural correlations are still overestimated, presumably due to the fact that we have completely neglected the flexibility of the individual monomers that allow for a larger variation of internal distances than assumed in the fjc model. On the other hand, the local correlation effects are even more pronounced for the polydisperse fc model. This results from the fact that we have to use the model also for the significant amount of small clusters, where the model is not applicable and local structural correlations are thus strongly overestimated. Therefore, we drop this model in the following. Nevertheless, we would like to point out that this deficiency could easily be overcome by using a base set of internal cluster structure factors derived from computer simulations of colloidal hard sphere clusters.
We also see from Fig.~\ref{fig:Sq-meas1} that our model systematically slightly underestimates the low-$q$ part of the structure factor. This is also due to the fact that we use an expression for monodisperse sticky hard spheres to calculate $S_{SHS}(0)$ and $S^{eff}_c(q)$. It is known that polydispersity not only decreases the amplitude of the nearest neighbor correlation peak but, for strongly correlated particles, it also results in an additional 'incoherent' contribution to the intensity and thus increases $S^{eff}_c(0)$~\cite{klein2002interacting}. Unfortunately, we have no simple analytical expression that would allow us to calculate the measured structure factor for our models in this case. 
However, given the many approximations made and the simple coarse-grained models used, we believe that the agreement between the experimental SAXS data and the calculated structure factors shown in Fig. \ref{fig:Sq-meas1} is quite remarkable, especially given that once we have fixed the parameters from our analysis of the SLS data there remain no additional free parameters to be adjusted. This is further illustrated in Fig.~\ref{fig:Sq-meas-all}, where we summarise the comparison between experimental data and calculated structure factors based on the polydisperse fjc model for both ionic strengths and all concentrations investigated. The agreement is indeed quite remarkable, and indicates that our model well captures both the concentration and ionic strength-dependent self-assembly into small clusters as well as the structural signature of these clusters in SAXS experiments.

\begin{figure}[bt!]
	\centering
	\includegraphics[width=0.6\linewidth]{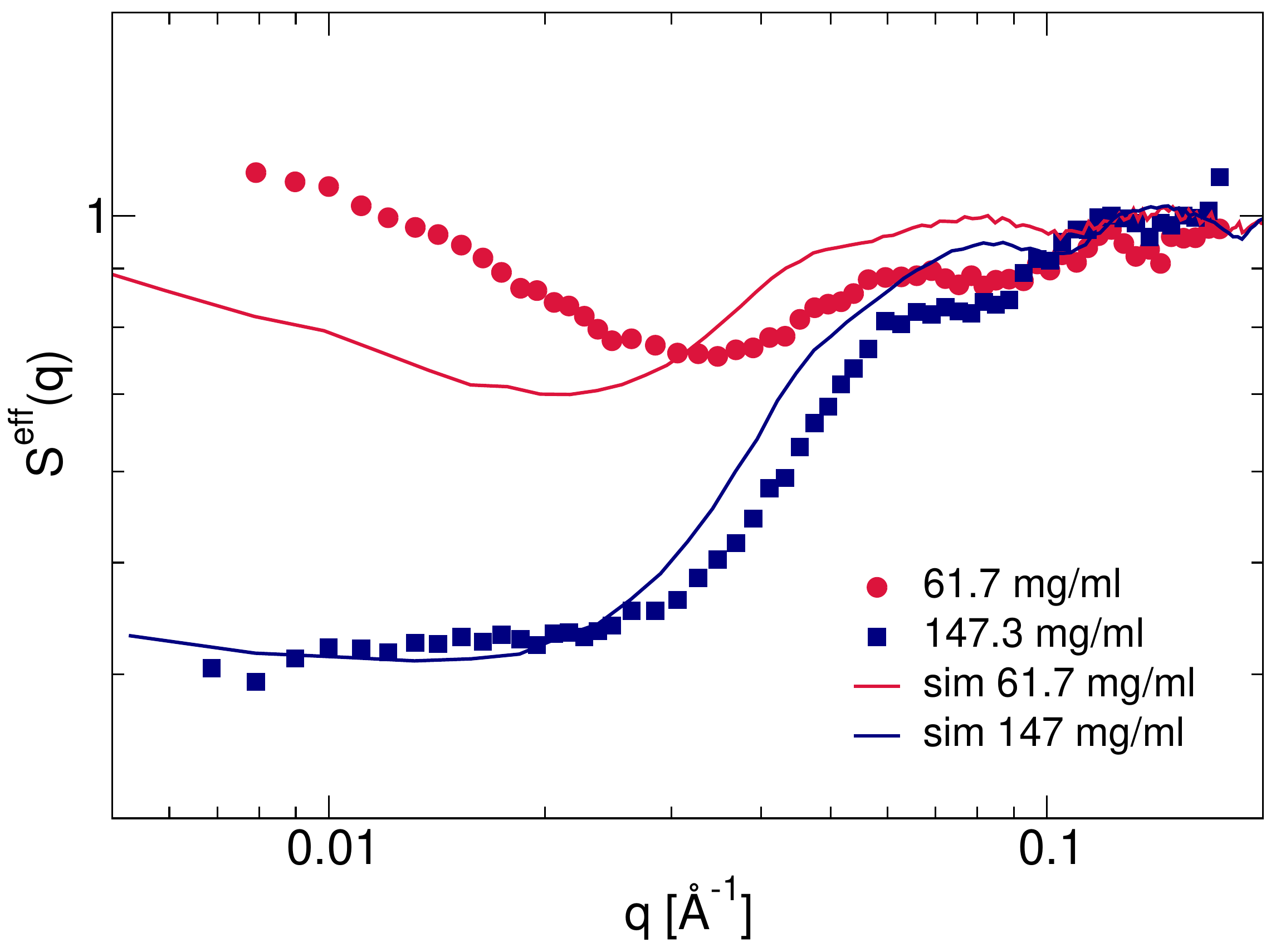}
	\caption{\footnotesize Measured effective structure factor $S^{eff}(q)$ compared with the corresponding one calculated from computer simulations (Eq.~\ref{Sqtot-sim}) at two mAb concentrations at 10 mM NaCl (red: 61.7 mg/ml; blue: 147.3 mg/ml). Symbols are measured experimental values, solid lines are data from MC simulations.}
	\label{fig:Sq-meas-MC}
\end{figure}

The full effective structure factor $S^{eff}(q)$ can also be obtained from computer simulations for the 9-bead model (Eq.~\ref{Sqtot-sim}). The results are shown in Fig.~\ref{fig:Sq-meas-MC}. For the highest concentration the agreement between simulations and experiments is fairly good, and both the low-$q$ values as well as the full $q$-dependence of the simulated structure factor $S^{eff}_{sim}(q)$ match the experimental $S^{eff}(0)$ as well as the measured $S^{eff}(q)$ almost quantitatively, indicating that for the chosen parameters our simple patchy 9-bead model reproduces the cluster size distribution and the structural correlations well.

At the lower concentration of $c = 61.7$ mg/ml, however, we do observe systematic deviations. While the low-$q$ values seem to reach the correct asymptotic $S^{eff}(0)$ value for this sample provided that we could extend the current $q$-range by going to a larger cell with more particles, there appears to be a systematic shift in the $q$-dependence between measured and simulated structure factors. We believe that this is primarily caused by the absence of electrostatic interactions in our simple patchy model as well as by small differences in the geometrical dimensions of the real mAb and the 9-bead model already visible in Fig. \ref{fig:model}(b). Since we use the measurable quantity $R_g$ to convert simulation units to actual dimensions in nm, this results in slight differences between the effective bond lengths in the clusters, which are given by the diagonal distance between oppositely charged groups or patches for the real bead and the 9-bead model. In turn, this likely leads to a mismatch for the $q$-dependence of the cluster structure factor $S_c(q)$, and thus for the total $S^{eff}_{sim}(q)$ when plotting the results in real units of $q$ and not in dimensionless normalized units $qd$. While at lower concentrations the cluster form or structure factor dominates the total scattering intensity, and thus small systematic deviations become easily visible, at the highest concentration the total intensity and thus $S^{eff}(q)$ is dominated by cluster-cluster interactions, and these small shifts in $S_c(q)$ become less visible.

\section{Conclusions}

In this work, we provided a microscopic viewpoint on solutions of self-associating antibodies. In particular, by complementing multi-technique experiments with analytical derivations and numerical simulations, we thoroughly characterized the formation of clusters and their structural properties. Our work built on the 
exploitation of polymer and colloid theories, which has proven to be particularly effective for this purpose. In particular, we employed the freely jointed chain and the fractal colloid cluster models to derive expressions for the structure factor of clusters of various sizes and at different mAb concentrations. The theoretical predictions were then validated by computer simulations, in which a rigid 9-bead model explicitly accounting for the anisotropic Y-shape of antibodies was used, and subsequently tested against experiments. While the freely joint chain model provides a sound description for a wide range of cluster sizes, the colloid model clearly appears to best suited only for clusters of intermediate and large dimensions, whose number of monomers is not less than 15 units. The excellent agreement between the different methodologies allowed us to provide a first microscopic characterization of mAb clusters. Specifically, we found that their structure is independent of the concentration of the antibody solution for a specific cluster size, and that the internal structure of clusters with few monomers is strongly dependent on their size at low scattering vectors. The solution structure factor, calculated with the fjc model, was then successfully compared to that obtained by SAXS experiments, demonstrating the validity of this model for the range of concentrations and ionic strengths investigated. In this way, having been able to decipher the contribution of individual clusters, we also got information on the collective response of the polydisperse cluster fluid as a function of concentration. As a result, our theoretical approach was favourably compared to microrheology data, being able to describe the dependence of the relative viscosity on mAb concentration over an extended range of concentrations and for two different ionic strengths. 

Our results thus provide a direct microscopic confirmation of the fact that the formation of small and medium-sized clusters is critical in the concentration-dependent increase of viscosity for this type of self-assembling antibodies. However, although our simplified model can provide important information at the qualitative level, a more faithful modeling of the molecule will have to be pursued in the future in order to reach a more quantitative assessment of the macroscopic response of this type of solutions. To this aim, two important aspects will need to be taken into account, namely the possible contributions from the internal flexibility of the antibody molecule and the effect of its inhomogeneous charge distribution. In the former case, while it is known that flexibility between domains of the antibody is crucial to the immunological response, it is not yet clear how relevant this is to the assembly of antibodies with attractive domains and their resulting solution structure. At the same time, a more refined treatment of charges beyond the patchy minimal model, which includes screening effects, may lead to a more thorough understanding of the mechanisms of assembly between antibodies in solution and the exact shape and arrangement of the clusters. The study presented here thus represents a first step for understanding the collective behavior of the solutions in terms of the individual elements that populate the system. This approach will also be instructive for other types of antibodies with different properties, with the final aim to improve the formulation of stable, low-viscosity solutions of therapeutic monoclonal antibodies. 

\section{Supporting Information}

Antibody excluded volume, simulation snapshots

\section{Acknowledgments}

We thank T. Garting for help with the microrheology measurements, and C. Rieschel for helpful discussions. This work was financed by the Swedish Research Council (VR; Grant No. 2016-03301, 2018-04627 and 2022-03142), the Faculty of Science at Lund University, the Knut and Alice Wallenberg Foundation (project grant KAW 2014.0052), the European Research Council (ERC-339678-COMPASS) and Novo Nordisk. The SAXS measurements at low concentrations were performed at the SWING beamline of the synchrotron SOLEIL, and we gratefully acknowledge the help of the local contact J. Perez, and the support from S. R. Midtgaard and A. Haahr Larsen.

\clearpage
\newpage

\large
{\centering
\section*{Using cluster theory to calculate the experimental structure factors of antibody solutions\\Supplementary Information}
\normalsize

\noindent  Nicholas Skar-Gislinge\textsuperscript{ 1,2}, Fabrizio Camerin\textsuperscript{ 3}, Anna Stradner\textsuperscript{ 1,4}, Emanuela Zaccarelli\textsuperscript{ 5,6}, Peter Schurtenberger\textsuperscript{ 1,4}\\
\medskip
\small
\textit{%
\textsuperscript{1}Physical Chemistry, Department of Chemistry, Lund University, SE-221 00 Lund, Sweden\\
\textsuperscript{2}Copenhagen Business School, Porcelaenshaven 18B, 2000 Frederiksberg, Denmark\\
\textsuperscript{3}Soft Condensed Matter, Debye Institute for Nanomaterials Science, Utrecht University, Princetonplein 5, 3584 CC Utrecht, The Netherlands\\
\textsuperscript{4}LINXS - Lund Institute of advanced Neutron and X-ray Science, Scheelevägen 19, SE-223 70 Lund, Sweden\\
\textsuperscript{5}Institute for Complex Systems, National Research Council (ISC-CNR), Piazzale Aldo Moro 5, 00185 Rome, Italy\\
\textsuperscript{6}Department of Physics, Sapienza University of Rome, Piazzale Aldo Moro 2, 00185 Rome, Italy
}
}

\normalsize
\renewcommand{\thefigure}{S\arabic{figure}}\setcounter{figure}{0}

\section{Antibody excluded volume}
\begin{figure}[h!]
\centering
\includegraphics[width=0.6\linewidth]{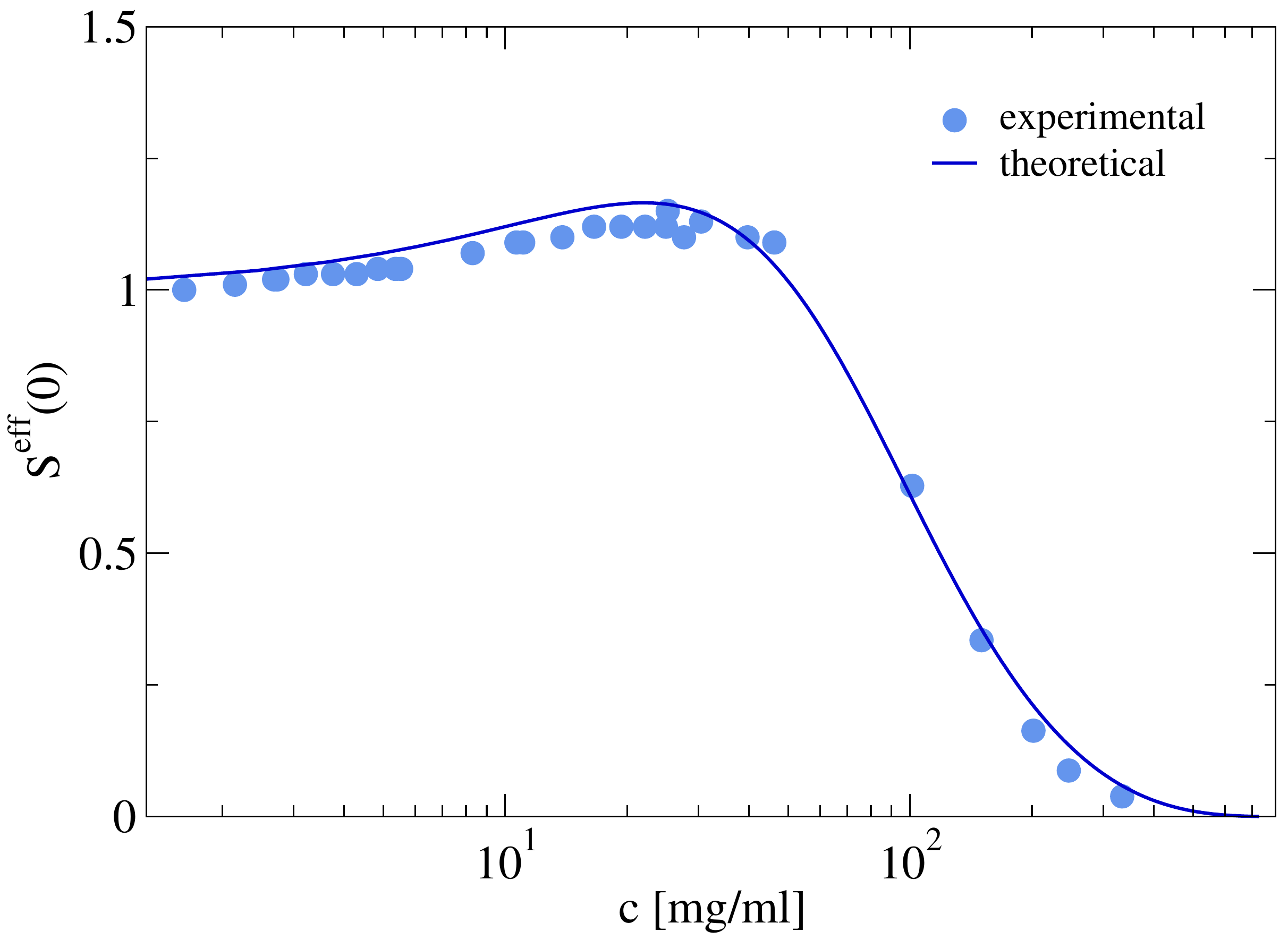}
 \caption{Experimental and theoretical $S^{eff}(0)$ ($\sigma_{HS}=2.95\sigma$, $T=0.11$) as a function of concentration.}
	\label{S0}
\end{figure}

For developing a correct theoretical and computational model for antibodies, it is appropriate to check whether its excluded volume is correct with respect to experimental evidence. To this aim, we exploit the relationship between isothermal compressibility and the static structure factor $S(q=0)$ via the number density $\rho$.

As explained in the main text, it is possible to map the coarse-grained Y-model to an effective hard sphere. To do this, as shown in Fig.~S1, we first match the experimentally determined $S^{eff}(0)$ as a function of the concentration with the same curve obtained via Wertheim theory, allowing to identify the correct size of an effective hard sphere diameter $\sigma_{HS}$.
Subsequently, to have a Y model that matches correctly the excluded volume, we compare the Y-simulation results of $S^{eff}(0)$ with the predictions of Carnahan-Starling obtained by using the $\sigma_{HS}$ previously determined, as reported in Fig.~S2. 
Here we compare the previously used 6-bead~\cite{Skar-Gislinge2019} and the newly chosen 9-bead model. We find that for the former case, the best agreement is obtained for $\sigma_{HS}=2.7\sigma$ at low density and for $\sigma_{HS}=2.54\sigma$ at higher concentrations. The latter value does not coincide with the one matching Wertheim theory results in the presence of patches~\cite{Skar-Gislinge2019}. 
Instead, for the 9-bead model, we find that the isothermal compressibility actually reproduces the correct behavior with the expected HS size, namely $\sigma_{HS}=2.95\sigma$. Importantly, the 9-bead model is able to capture the hard sphere compressibility in the whole experimental concentration range, thus appearing to be a superior model than the 6-bead model. Therefore, despite a slight increase in the number of beads, in the main text we focus on this model to provide a comparison for the experimental structure factors.

\begin{figure}[b!]
\centering
\includegraphics[width=0.6\linewidth]{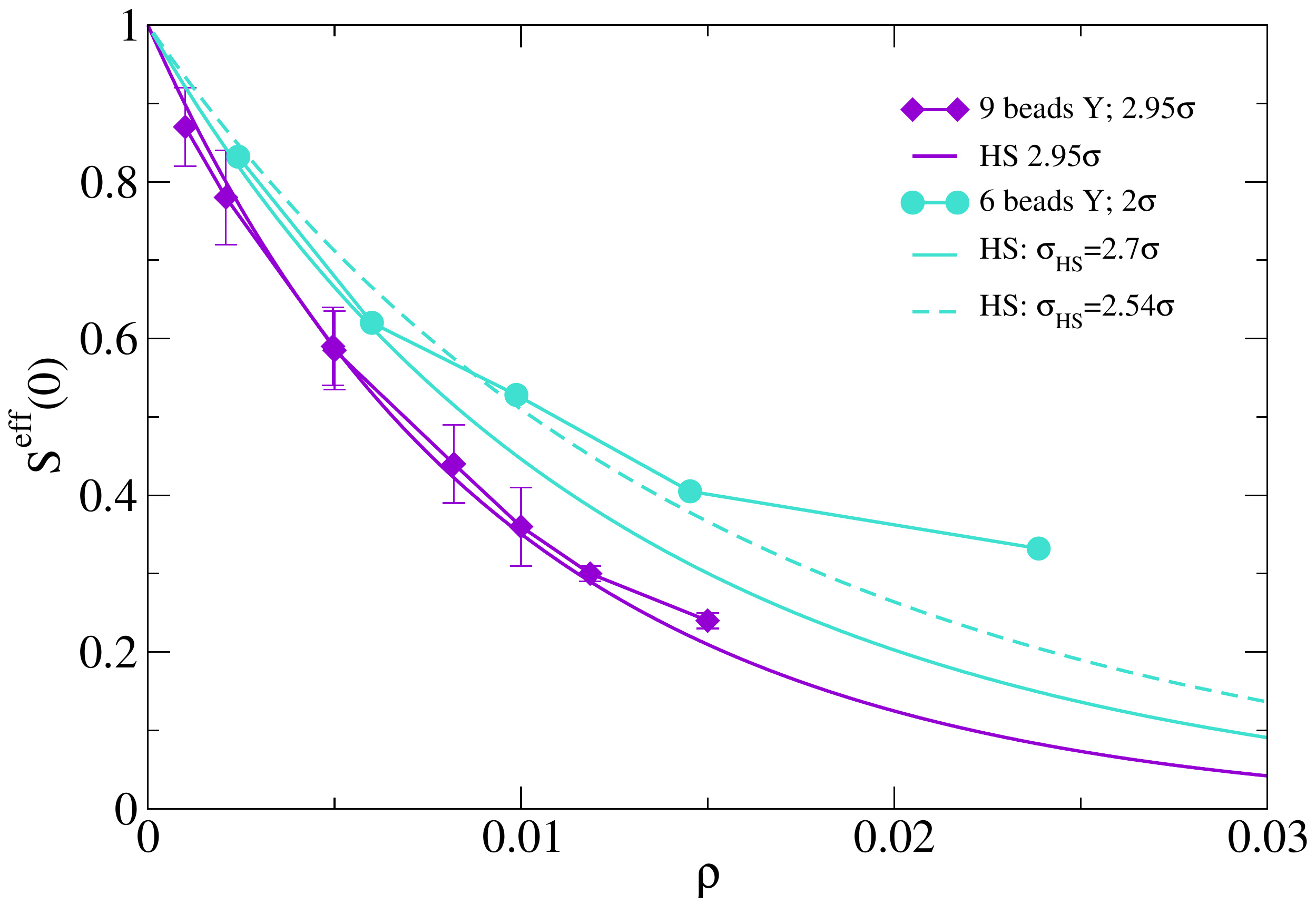}
 \caption{$S^{eff}(0)$ as a function of the number density $\rho$ for $9$ and $6$-beads Y models. The corresponding Carnahan-Starling (CS) results are also reported.}
	\label{S0}
\end{figure}

\section{Simulation snapshots}

\begin{figure}[H]
\centering
\includegraphics[width=0.9\linewidth]{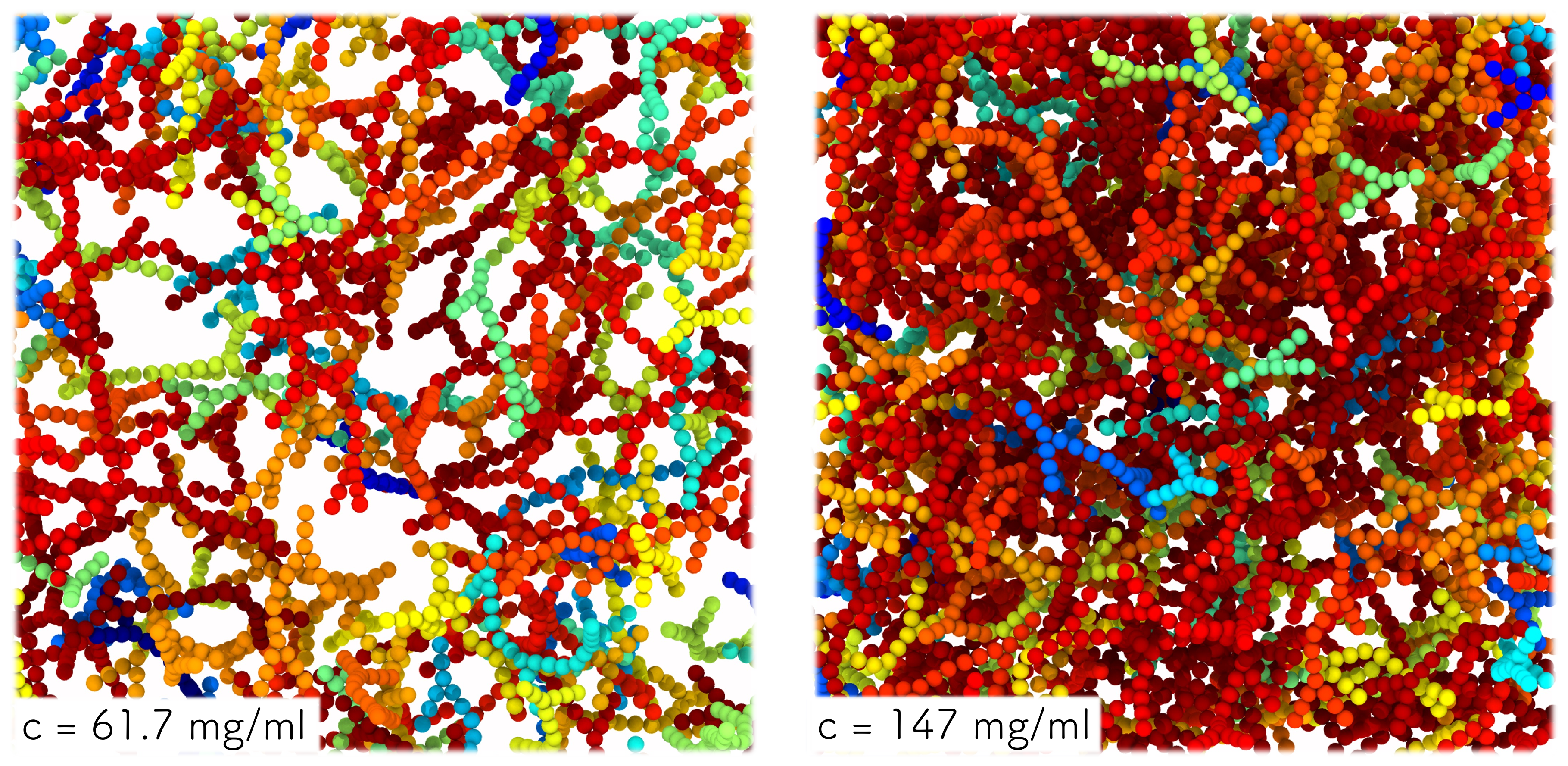}
 \caption{Simulation snapshots of the antibody system for $c=61.7$ and $147$ mg/ml. Clusters are highlighted with different colors. Antibodies belonging to the same cluster are colored alike.}
	\label{snapshots}
\end{figure}

In Figure S3, we report two simulation snapshots for two different concentrations analyzed in the main text, namely $c=61.7$ mg/ml and $147$ mg/ml. Individual clusters of different sizes are reported in the main text.

\bibliography{mAb-bib}

\end{document}